\documentclass[letterpaper]{IEEEtran}
\ifCLASSINFOpdf
\else
\usepackage[dvips]{graphicx}
\fi
\usepackage[cmex10]{amsmath}
\usepackage{upgreek}
\usepackage{booktabs}
\usepackage{geometry}
\usepackage{epsfig}
\usepackage{latexsym}
\usepackage{multirow}
\usepackage{stfloats}

\usepackage{epstopdf}
\usepackage{color}  
\usepackage{tabularx} 
\usepackage{amssymb}
\usepackage{enumerate}
\graphicspath{{./Figures/}}
\usepackage{color}
\usepackage{bbm}
\usepackage{bm}
\usepackage{cite}
\usepackage{subfigure}
\usepackage{balance}
\usepackage{mathrsfs}
\usepackage{verbatim}
\usepackage{dsfont}
\usepackage{verbatim}
\usepackage{makecell}
\usepackage{algorithm}
\usepackage{algorithmic}
\usepackage{tikz}
\usepackage{diagbox}
\usepackage{caption}
\usepackage[framemethod=tikz]{mdframed}
\usepackage{multicol}
\usepackage{environ}
\usepackage{tikz}
\begin{document}
\title{Frequency-Position-Fluid Antenna Array and Beamforming for Ultra-dense Connectivity in Terahertz Wireless Systems}
\author{Heyin~Shen, Chong~Han,~\IEEEmembership{Senior Member,~IEEE,}
and~Jinhong~Yuan,~\IEEEmembership{Fellow,~IEEE}
\thanks{Heyin Shen and Chong Han are with the Terahertz Wireless Communications (TWC) Laboratory, Shanghai Jiao Tong University, Shanghai 200240, China (e-mail: heyin.shen@sjtu.edu.cn; chong.han@sjtu.edu.cn).}
\thanks{Jinhong Yuan is with the School of Electrical Engineering and Telecommunications, University of New South Wales, Sydney, NSW 2052, Australia (e-mail: j.yuan@unsw.edu.au).}
}

\markboth{}
\MakeLowercase
\maketitle
\begin{abstract}
\boldmath
To support ultra-dense connectivity in terahertz (THz) communications, this paper proposes a dynamic frequency-position-fluid antenna (D-FPFA) architecture. Frequency-tunable local oscillators (LOs) are integrated into the RF chains to access different sub-bands, thereby expanding the total bandwidth of the system and providing frequency-domain diversity. To exploit spatial diversity, the base station employs movable subarrays, and each user is equipped with a movable antenna. We first develop a two-phase beam-split-aware frequency allocation strategy. In the first phase, we divide users into disjoint sub-bands according to their channel correlation coefficients to mitigate the interference. In the second phase, we investigate the wideband near-field beam-split effect for planar arrays and reveal an astigmatism phenomenon, in which the beam at a non-central subcarrier cannot be perfectly refocused at a single spatial point. Then, we establish a beam split multiplexing strategy, where we formulate the user grouping task as a minimum dominating set problem. To maximize the sum rate, we introduce a switch network along with a distance-based antenna selection strategy to account for the near-field channel gain variations, followed by a particle swarm optimization-based algorithm that jointly optimizes the antenna positions and precoders. Numerical results show that, the proposed D-FPFA achieves approximately $2.3$ times the sum rate of a conventional phase-shifter (PS)-based array-of-subarrays (AoSA) architecture. It also attains 95\% of the sum rate of its TTD counterpart while providing approximately $2.8$ times its energy efficiency (EE). Moreover, the fully connected variant of D-FPFA, i.e., FPFA, achieves the highest EE among all considered architectures. 
\end{abstract}
\IEEEpeerreviewmaketitle
\section{Introduction}
\label{section_intro}
Driven by the availability of multi-GHz continuous bandwidth, Terahertz (THz) wireless communication has emerged as a compelling solution for realizing terabit-per-second (Tbps) data rates in next-generation networks~\cite{Challenges,above100G}. To boost spectral efficiency (SE), multiple-input multiple-output (MIMO) antenna architectures have been extensively researched, aiming to provide substantial array gain and mitigate the inherent challenge of limited propagation distance~\cite{dynamic,InterIntra}. However, a critical bottleneck resides in the reliance of conventional MIMO architectures which consist of numerous fixed-position antennas with rigid half-wavelength spacing. Such static configurations fail to exploit the full potential of spatial diversity offered by the continuous spatial variations of wireless channels, thereby severely constraining the available multiplexing performance~\cite{modeling}. This limitation becomes particularly acute under the 6G vision of ubiquitous connectivity, which targets a massive device density of $10^7$ devices/km$^2$~\cite{opportunistic}. Meeting such ultra-dense connection requirements necessitates significantly higher spatial degrees-of-freedom (SDoFs), creating an urgent need for more flexible antenna solutions.

To address the aforementioned challenges, position-fluid-antenna (PFA), or movable-antenna (MA)-aided systems, have been introduced~\cite{historical}. By utilizing mechanical drivers or conductive fluids, these antennas can dynamically adjust their physical locations within a designated region. This spatial flexibility allows the system to identify and occupy positions with optimal channel conditions, thereby enhancing performance~\cite{opportunities}. Extensive research has explored these technologies in wireless communications. In~\cite{movable-statistical}, the authors jointly optimize the beamforming matrices and the movable antenna selection. In~\cite{MA_graph}, the authors propose a graph theory-based algorithm to optimize the positions of the antennas. In multi-user scenarios, the authors in~\cite{multiple-access} have developed the fluid antenna multiple access scheme to exploit deep fades in inter-user interference, thus improving the signal-to-interference-plus-noise ratio (SINR). In~\cite{fluid-III}, a channel model for a multi-user PFA system has been developed under RIS-aided rich scattering environment, while in~\cite{movable_antenna1}, ZF-based and MMSE-based joint optimization of port selection and beamforming algorithms have been developed to minimize the transmit power. In~\cite{MU_uplink}, an uplink multi-user MA-enabled BS is considered, where the antenna positions, receive combining, and user transmit powers are jointly optimized to maximize the minimum achievable rate. Besides position and beamforming optimization, efficient channel acquisition over the movement region is also important for practical MA systems. In~\cite{MA_Globally_optimal}, the authors consider the discrete movement of MAs under both perfect and imperfect CSI conditions. In~\cite{MA-estimation}, a low-overhead compressed-sensing framework was developed to estimate the dominant multipath components from a limited number of measurements and then reconstruct the channel over the transmit and receive movement regions. However, most of the studies assume that all the antennas at the transmitter are movable, which will cause significant hardware complexity and power consumption. To solve this problem, a movable subarray configuration is studied in~\cite{MA_subarray} to balance performance and power consumption. Moreover, in~\cite{MA_subarray_1}, subarray-level movement is combined with hybrid beamforming for multiuser and multi-stream MIMO systems, while in~\cite{MA_subarray_MU}, an extremely large-scale movable-subarray architecture is proposed, where the long-term placement of multiple movable subarrays is optimized according to statistical channel information and user distributions. In practical MA systems, the aperture size is typically large to accommodate the antenna movement. This large aperture, together with the high frequencies used in the THz band, significantly extends the Rayleigh distance, thereby expanding the near-field communication range~\cite{NF_MA1}. Within this near-field region, the spherical wavefronts introduce significant spatial disparity in the channel gains of individual antenna elements relative to a target receiver. Consequently, antennas positioned farther from the receiver contribute substantially less to the overall channel strength while consuming the same amount of power as those in closer proximity~\cite{NF_effect}. This disparity presents a key spatial-domain challenge for the near-field MA systems.

Beyond the spatial domain, achieving ultra-dense connectivity also requires exploiting the abundant bandwidth in the THz band. With multi-GHz bandwidth, the frequency multiplexing technology can be utilized where the bandwidth is divided into sub-bands and allocated to different user equipment. As a result, the inter-band users can avoid interference by filtering out unwanted carriers. The authors in \cite{cluster-based} proposed a cluster-based multi-carrier bandwidth division multiple access scheme where spatially proximate users are served by different sub-carriers. In~\cite{SS-OFDMA}, a spatial-spread orthogonal frequency division multiple access has been developed where THzPrism beams are utilized to direct beams of different frequencies to distinct angles, thus covering users within a wide angular range. Considering the beam split effect which causes frequency-dependent beam directions or focal locations, extensive studies have been conducted on its characterization and compensation. For a near-field wideband ULA, closed-form expressions for the shifted angle and distance are derived, such that each subcarrier can be associated with one shifted focal point~\cite{rainbow}. Moreover, authors in~\cite{UCA} investigate the uniform circular array (UCA). It has been shown that exact focusing in UCA occurs only at the central frequency, while the perfectly phase-matched focal point for non-central subcarriers does not exist. However, the beam-split behavior of general two-dimensional planar arrays, such as UPAs, has not been fully characterized. Existing beamforming studies considering the beam-split effect mainly follow two directions. Most studies mitigate the beam split using true-time-delay (TTD) devices~\cite{EE_FTTD,DPP,D-TTD,TTD_config}. In contrast, other works exploit beam split for wideband THz transmission. In particular, the authors in~\cite{Beam_split_AOSA,Beam_split_MU} introduced a beam split multiplexing strategy for ULA systems, where users located within the beam split coverage can be served by different subcarriers of a single wideband beam and share one RF chain. Nevertheless, these existing studies mainly focus on specific array geometries, such as ULA or UCA, and do not characterize the beam-split behavior of a general two-dimensional planar aperture, such as the UPA system considered in this work.

Despite the vast bandwidth potential of THz frequencies, practical utilization is severely bottlenecked by hardware constraints—particularly the limited analog bandwidth of digital-to-analog converters (DACs). Conventional DACs exhibit analog passbands below 5 GHz\cite{above100G}. Consequently, a single THz RF chain can only access a fractional sub-band of around 5 GHz at baseband. To solve this problem, the tunable-frequency LOs can be utilized and are often implemented using voltage-controlled oscillators (VCOs), allowing for dynamic frequency adjustment via a control voltage. Current tunable-frequency LO technologies can be realized through oscillators based on III-V semiconductors~\cite{LO_3}, silicon-based solutions~\cite{LO_1,LO_2}, and schemes employing lower-frequency LOs followed by frequency multipliers to generate the desired high-frequency signal~\cite{LO_4,LO_5,LO_6}.

In light of these challenges and opportunities, we propose a dynamic frequency-position-fluid antenna (D-FPFA) architecture for wideband communications, which can effectively harvest multiplexing gains from both the spatial and frequency domains. Based on this architecture, the attainable sum rate is maximized to support ultra-dense connectivity through the joint design of frequency allocation, array configuration, and hybrid beamforming. The distinctive contributions of this work are summarized as follows.
\begin{itemize}
  \item We propose the D-FPFA architecture, which integrates tunable LOs to overcome the bandwidth constraints of limited DACs/ADCs. By dynamically shifting baseband signals across disjoint THz sub-bands, the system effectively realizes wideband coverage. To strike a balance between spatial flexibility and hardware cost, we adopt the movable subarray architecture at the BS, where each subarray moves as a unit within a predefined region. Complementing this transmitter design, we deploy fluid antennas at each user to construct a dual-sided movable antenna system, which enables the joint optimization of transceiver antenna positions. Moreover, by considering the NF effect where channel gain varies substantially across the array aperture, a switch network is inserted between the RF chains and the antenna array to enable dynamic antenna selection, which further improves the power utilization efficiency.

  \item To utilize the frequency resources enabled by the tunable LOs, we develop a two-phase frequency-allocation strategy. In Phase I, users with strong potential interference are assigned to disjoint sub-bands according to their channel correlation coefficients. In Phase II, the beam split within each sub-band is exploited to serve different users through different subcarriers of one wideband beam. Specifically, we reveal the astigmatism phenomenon in a near-field planar array that the beam at a non-central subcarrier cannot be perfectly refocused at one spatial point. Based on this analysis, we characterize the beam-split coverage using a gain-based condition and formulate the user-grouping problem as a directed minimum dominating set (MDS) problem.
  
  \item Based on the established user-frequency association, we further optimize the spatial-domain design of D-FPFA. Unlike conventional designs, the movable subarray in D-FPFA is shared by all frequency resources, such that it must simultaneously support all assigned users over the selected subcarriers. Therefore, we formulate a multi-band sum rate maximization problem over all user–frequency links. Moreover, to improve energy efficiency, we develop a distance-based dynamic antenna-selection method using a switch network, which prioritizes antenna elements with stronger channel gains to reduce energy waste. With the active antenna connections determined, a PSO-based algorithm is then developed to jointly optimize the transceiver antenna positions and hybrid precoders.

  \item Comprehensive simulations are conducted to evaluate the performance of the proposed D-FPFA architecture. The results show that D-FPFA achieves approximately $2.3$ times the sum rate of a conventional array-of-subarrays (AoSA) architecture. It also attains 95\% of the sum rate of its ideal-TTD counterpart while providing approximately $2.8$ times the EE. While the fully connected variant of the D-FPFA, i.e., FPFA, achieves the highest EE among all considered architectures. Overall, D-FPFA achieves the highest sum rate among practical PS-based designs and maintains competitive EE, whereas FPFA achieves the highest EE among all considered architectures and provides the best rate-power balance.
\end{itemize}

The remainder of this paper is organized as follows. In Sec.~\ref{section_Sys_channel_Model_wideband}, we present the wideband channel model and the system model for D-FPFA. Then, in Sec.~\ref{Sec_two_phase_frequency}, we propose a two-phase frequency allocation strategy to allocate the subcarriers to each user through wideband beam split multiplexing. In Sec.~\ref{sec_problem_P3}, we explore the spatial domain diversity where we formulate the optimization problem to maximize the sum rate. Then, we design the dynamic antenna selection strategy. After that, we jointly optimize the precoders and the MA positions using a PSO-based algorithm. Simulation results are shown in Sec.~\ref{section_performance} where we evaluate the performance of the proposed algorithms with different architectures. Conclusions are drawn in Sec.~\ref{section_conclusion}.

\textit{Notations:} $\mathbf{A}$ is a matrix, $\mathbf{a}$ is a vector and $a$ is a scalar. $(\cdot)^*$ denotes the complex conjugate of a vector; $(\cdot)^T$ and $(\cdot)^H$ represent the transpose and conjugate transpose of a matrix. $| \cdot |$,  $\Vert \cdot \Vert_2 $ and $\Vert \cdot \Vert_F$ denote the modulus, the Euclidean norm and the Frobenius norm. $\operatorname{det} (\cdot)$ represents the determinant. $\mathbb{C}^{m\times n}$ is the set of complex-valued matrices of dimension $m \times n$. $\odot$ denotes the element-wise multiplication of two matrices.

\begin{figure*}
    \centering
        \includegraphics[width = 0.9\textwidth]{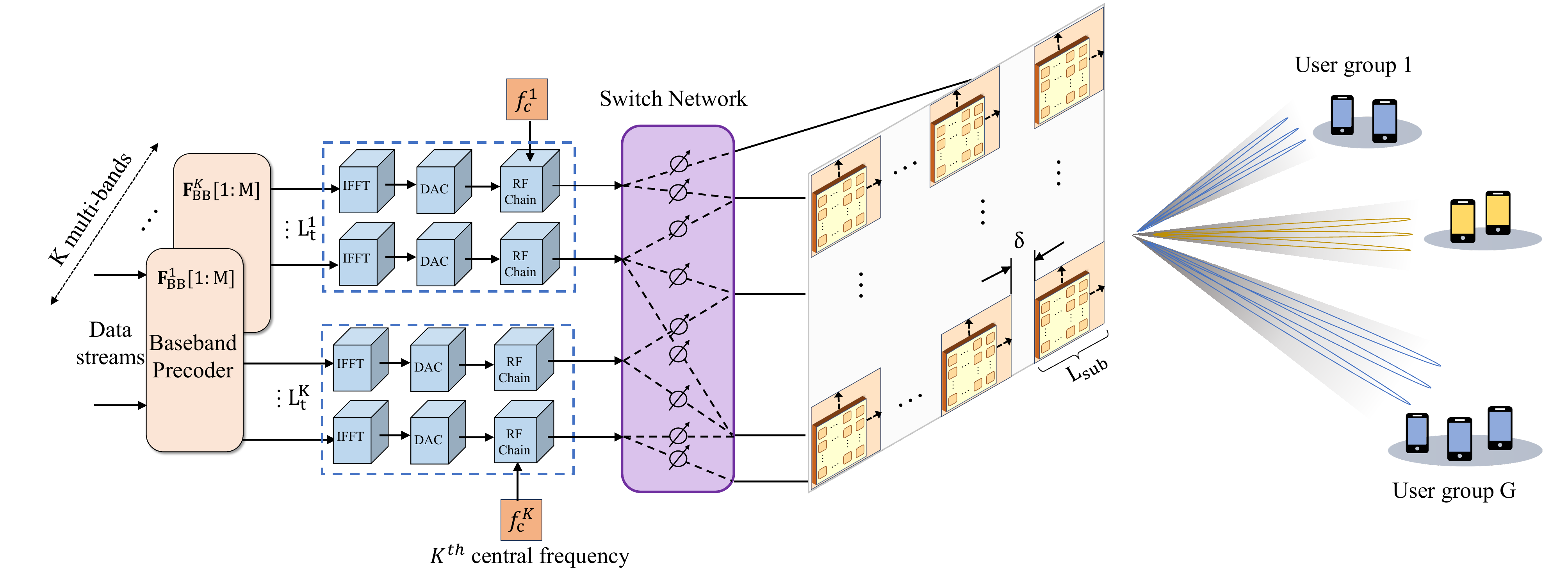}
    \caption{Block diagram of the THz multi-user D-FPFA hybrid beamforming system.}
    \label{Fig.model}
\end{figure*}

\section{System and Channel Models}
\label{section_Sys_channel_Model_wideband}
In this section, we introduce the channel model and the system model for the D-FPFA hybrid beamforming architecture. As shown in Fig.~\ref{Fig.model}, we consider a downlink wideband communication system with ultra-dense connectivity, where a BS with $N_t$ antennas and $L_t$ RF chains serves $U$ single-MA users simultaneously with $U > L_t$. At the BS, the transmitted signal is divided into $K$ sub-bands with equal bandwidth $B$, where each sub-band consists of $M$ subcarriers. Therefore, the available bandwidth of the whole system increases to $KB$ and the total number of carriers is denoted as $L = MK$. Each RF chain is equipped with a frequency-tunable LO. Within a transmission block, each tunable LO selects one central frequency from $\{f_c^1, f_c^2, \ldots, f_c^K\}$, and the corresponding RF chain generates one wideband beam over the selected sub-band. The central frequency of an RF chain can be reconfigured across different transmission blocks via the tunable LO, but each RF chain operates at only one central frequency within a given block. The $m^{\rm th}$ subcarrier at central frequency $f_c^k$ is denoted as $f_m^k = f_c^k+\frac{B}{M}(m-1-\frac{M-1}{2}), m = 1,\cdots, M$. We write the number of RF chains at central frequency $f_c^k$ as $L_t^k$ such that the total number of RF chains is expressed as $L_t = \sum_{k=1}^K L_t^k$. 
At the receiver side, the massive users are separated into $G = L_t$ user groups. Each group is served exclusively by one wideband beam generated by one RF chain where the signals towards different users within one group are modulated on different subcarriers. As a result, interference within each group can be eliminated by filtering out the unwanted subcarriers. 
Note that when $f_c^1 = f_c^2 = ... = f_c^K$, the system degrades to a traditional frequency-fixed wideband system. 

To enhance SDoF, $N_{sub}$ movable subarrays are equipped to form a uniform planar array (UPA). We define the bottom-left antenna of each movable subarray as its reference antenna and allow each reference antenna to move freely within a square region of side length $p\lambda_{\text{max}}$. Therefore, the side length of the movable region of each subarray can be calculated as
\begin{equation}
    L_{sub} = \left(p+\frac{\sqrt{N_{t}^{sub}}-1}{2}\right)\lambda_{\text{max}},
\end{equation}
where $N_{t}^{sub} = N_t/N_{sub}$ denotes the number of transmit antennas within one movable subarray. To avoid antenna coupling effect and subarray overlap during movement, adjacent movable regions should be separated by at least $\delta \geq\frac{1}{2}\lambda_{\text{max}}$, where $\lambda_{\text{max}} = \frac{c}{f_c^{\text{min}}}$. As a result, the movable subarrays form a non-uniform array configuration, which enlarges the overall array aperture, thus leading to the near-field communications. Due to the severe path loss in the THz band, we focus on the line-of-sight (LoS) case. Therefore, the near-field LoS wideband channel model for a user $u$ at subcarrier $f_m^k$ can be written as
\begin{equation}
\label{eq:channel}
\begin{aligned}
    \mathbf{h}_u^k[m] =  & [|\alpha_{u1}[m]| e^{-j\frac{2\pi f^k_m}{c}\lVert \mathbf{p}_t^1-\mathbf{p}_u\rVert_2}, \cdots, \\
    & |\alpha_{uN_t}[m]| e^{-j\frac{2\pi f^k_m}{c}\lVert \mathbf{p}_t^{N_t}-\mathbf{p}_u\rVert_2}],
\end{aligned}
\end{equation}
where $\mathbf{p}_t^i$ and $\mathbf{p}_u$ denote the position of the $i^{\mathrm{th}}$ transmit antenna and the position of the $u^{\rm th}$ user, respectively. $|\alpha_{ui}[m]|$ represents the amplitude of the complex path gain which is modeled by considering both the free-space path loss (FSPL) and the molecular absorption loss in the THz band. Specifically, denote the transmit distance as $d_{u,i}=\lVert\mathbf p_i^t-\mathbf p_u\rVert_2$. Then $|\alpha_{u,i}[m]|=\frac{c}{4\pi f_m^k d_{u,i}}
\exp\left(-\frac{1}{2}\kappa(f_m^k)d_{u,i}\right)$,
where $\kappa(f_m^k)$ is the molecular absorption coefficient at frequency $f_m^k$. The near-field channel response vector can then be written as 
\begin{equation}
\begin{aligned}
    & \mathbf{a}(f_m^k,\mathbf{p}_u,\{\mathbf{p}_n^t\}) \\
    = & [e^{-j\frac{2\pi f^k_m}{c}\lVert \mathbf{p}_t^1-\mathbf{p}_u\rVert_2},\cdots, e^{-j\frac{2\pi f^k_m}{c}\lVert \mathbf{p}_t^{N_t}-\mathbf{p}_u\rVert_2}].
\end{aligned}
\end{equation}

Due to the large array aperture and the wide frequency bands, signals transmitted from different antennas to a target user will experience significant variations in path loss. However, in FC architectures, antennas with different contributions consume the same amount of power and require the same number of phase shifters, resulting in energy inefficiency. Therefore, a switch network is added to enable dynamic control of the number of antennas connected to each RF chain. Denote $\mathbf{S}^k \in \mathbb{C}^{N_t \times L_t^k}$ as the switch matrix. Then, it must satisfy $\mathbf{S}^k[i,j] \in \{0,1\}$ where $0$ indicates no connection between the $i_{\rm th}$ antenna and the $j^{\rm th}$ RF chain while $1$ represents an active connection. Denote $\mathbf{F}^k_{\rm RF} \in \mathbb{C}^{N_t \times L_t^k}$ as the phase shifter matrix at central frequency $f_c^k$ which satisfies $\mathbf{F}^k_{\rm RF}[i,j] = 1, \forall i,j$. Then, the overall analog precoder is written as
\begin{equation}
    \widetilde{\mathbf{F}}^k_{\rm RF} = \mathbf{S}^k \odot \mathbf{F}^k_{\rm RF}.
\end{equation}
Denote $\mathcal{U}_m^k$ as the set of users sharing the subcarrier $f_m^k$, then the received signal can be represented as
\begin{equation}
\begin{aligned}
    \mathbf{y}_{u}^k[m] & = \mathbf{h}_{u}^k[m]\widetilde{\mathbf{F}}^k_{\rm RF}\mathbf{F}_{\mathrm{BB}}^k[m]\mathbf{s}^k[m]+\mathbf{n}_{u}^k[m]\\
    & = \underbrace{\mathbf{h}^k_{u}[m]\widetilde{\mathbf{F}}^k_{\rm RF}\mathbf{f}^k_{\mathrm{BB},u}[m]\mathbf{s}^k_u[m]}_{\text {desired signal}} \\
    & +\underbrace{\mathbf{h}^k_{u}[m]\widetilde{\mathbf{F}}_{\rm RF}^k \sum_{u^{\prime}\in \mathcal{U}_m^k} \mathbf{f}^k_{\mathrm{BB},u^{\prime}}[m] s^k_{u^{\prime}}[m]}_{\text{inter-group interference}}+\mathbf{n}_{u}^k[m],
\end{aligned}
\end{equation}
where $\mathbf{s}^k[m] = [\mathbf{s}^k_{u_1}[m], \mathbf{s}^k_{u_2}[m],\cdots]^T$ and $\mathbf{F}_{\mathrm{BB}}^k[m] = [\mathbf{f}^k_{\rm BB,u_1}[m], \mathbf{f}^k_{\rm BB,u_2}[m],\cdots], u_i \in \mathcal{U}_m^k$ denote the transmitted symbols and the digital precoder, respectively. $\mathbf{n}_{u}^k[m]$ denotes the i.i.d complex Gaussian noise, i.e., following $\mathcal{CN}(0,\sigma_n^2)$. Thus, the SINR for each user can be written as
\begin{equation}
\label{eq:SINR}
    \gamma_u^k[m]=\frac{p_u^k[m]^2\left|\mathbf{h}_u^k[m] \widetilde{\mathbf{F}}_{\rm RF}^k \mathbf{f}_{\mathrm{BB}, u}^k[m]\right|^2}{\sum_{i\in \mathcal{U}^k_m, i \neq u} p_i^k[m]^2\left|\mathbf{h}_u^k[m] \mathbf{F}^k_{\rm RF} \mathbf{f}^k_{\mathrm{BB}, i}[m]\right|^2+\sigma_n^2},
\end{equation}
where $p_u^k[m]$ denotes the power allocated to the $u^{\mathrm{th}}$ user at subcarrier $f_m^k$. For simplicity, we assume equal power allocation in this work, i.e., $\sum_{(k,m)\in \mathcal{M}_u}p_u^k[m]=P_t/U$ where $\mathcal{M}_u$ denotes the set of subcarriers assigned to user $u$. As a result, the SE for each user at frequency $f_m^k$ is presented as
\begin{equation}
    SE_u^k[m] = \log_2(1+\gamma_u^k[m]).
\end{equation}

\section{Two-phase Frequency Allocation}
\label{Sec_two_phase_frequency}
In the considered massive access scenario, users are densely deployed and often located in close spatial proximity. Relying solely on the spatial domain to distinguish these users is challenging due to hardware limitations, such as the limited number of RF chains and the physical beamwidth. When the number of users exceeds the spatial multiplexing capability, the BS cannot generate sufficient independent beams to separate them. This inevitably results in severe inter-user interference, which drastically degrades the system performance. To address this issue and minimize interference, we propose a two-phase hierarchical frequency allocation strategy. In the first phase, the $K$ sub-bands are allocated, and in the second phase, distinct subcarriers within each sub-band are assigned to individual users covered by the same wideband beam.

\subsection{Phase I: Sub-band Allocation}
\label{Sexc_frequency-allocation}
We first consider the sub-band allocation phase where users with high interference are divided into different sub-bands. As directly computing the interference requires the knowledge of the beamforming matrix, we apply the sum channel correlation coefficient as a measure of the severity of the interference \cite{NF_correlation1}. In this work, the channel vectors used to calculate the correlation coefficient are obtained from the near-field SWM defined in Eq.~(2). For simplicity, we temporarily omit the subcarrier index $m$ such that the correlation coefficient of two users in the same sub-band $k$ can be expressed as
\begin{equation}
    \rho_{ij}^k = \frac{|\mathbf{h}_i^k(\mathbf{h}^k_j)^H|}{\Vert\mathbf{h}_i^k\Vert \Vert\mathbf{h}_j^k\Vert}.
\end{equation}
Note that a larger $\rho_{ij}^{k}$ indicates a stronger potential inter-user interference~\cite{NF_correlation1}.
Therefore, the optimization problem is formulated as
\begin{subequations}
\begin{align}
\mathcal{P}_1: \underset{\{\mathcal{U}_k\}} {\operatorname{minimize}} & \sum_{k=1}^K \sum_{\substack{i,j \in \mathcal{U}_k \\ i \neq j}} \rho_{ij}^k, \\
    \text{s.t.}  \quad \ \ 
    & \{1,\cdots,U\} = \bigcup_{k=1}^{K}\mathcal{U}_{k},
    \label{C1}\\
    & \mathcal{U}_{k}\cap\mathcal{U}_{k'} = \varnothing, k\neq k'. \label{C2}
\end{align}
\end{subequations}
Constraint~\eqref{C1} indicates that all $U$ users are covered by the $K$ sub-bands, and~\eqref{C2} indicates that each user can only be in one sub-band. 
To solve this problem, we first assume all users are in the same sub-band. Then we find the user pairs with the largest $\rho_{ij}$ and remove one of them from the current sub-band. Next, we search the $K$ sub-bands to find a new group for this user based on the criterion that the resulting sum interference is the smallest. Therefore, the sub-band allocation algorithm can be summarized as follows:
\begin{itemize}
\item Step 1: Calculate the correlation coefficient for each user pair and sort them in descending order as $\{ \rho_1, \rho_2,...,\rho_i,...\rho_{U(U-1)/2}\}$. Set $i = 1$.
\item Step 2: Find the user pairs with coefficient $\rho_i$. If they are in the same group, remove one of the users $u$ out of the current group. 
\item Step 3: Select a new group such that the sum correlation coefficient is the smallest after adding $u$ to the group. Set $i = i+1$, and go back to Step 2 until $i = U(U-1)/2+1$.
\end{itemize}

\begin{figure}
    \centering
    \includegraphics[width=0.48\textwidth]{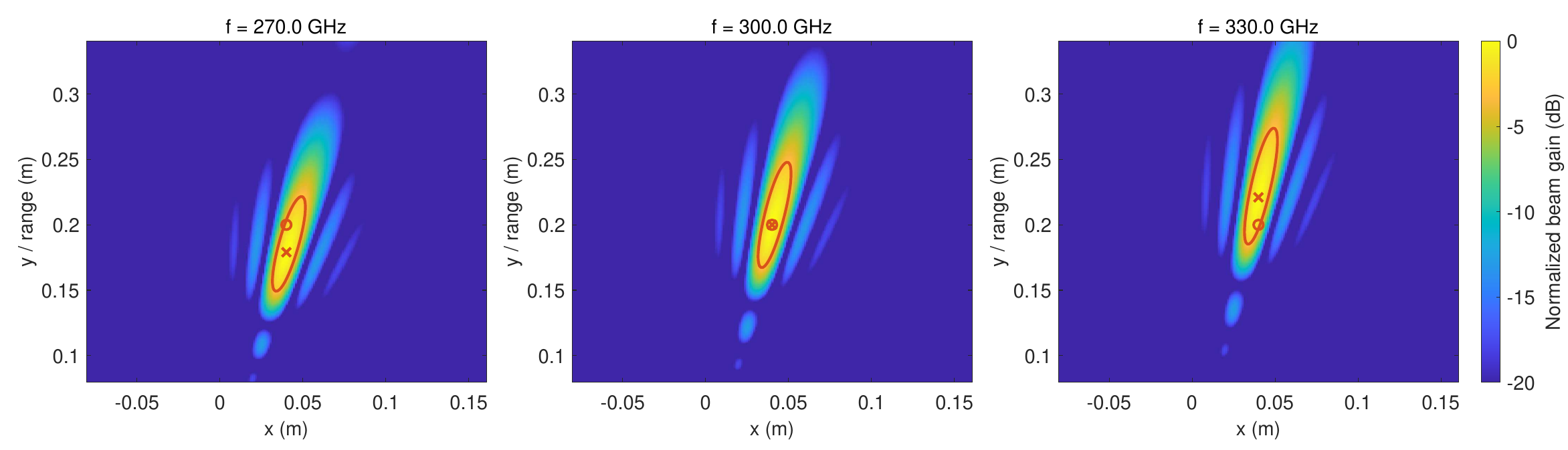}
    \caption{Beam pattern for wideband subcarriers in near-field UPA.}
    \label{fig:3dB_coverage}
\end{figure}

\subsection{Phase II: Beam Split Multiplexing}
\label{Sec_beam_split}
After allocating the $K$ sub-bands, we further group users within each sub-band. Specifically, users located within one wideband beam are assigned different subcarriers to reduce the number of RF chains and inter-user interference. To characterize this beam-split coverage relation, we first consider an arbitrary candidate user $u_i\in\mathcal U_k$, located at $\mathbf p_i$, toward which the central-frequency analog beam is focused. We refer to $u_i$ as the group center because its focused wideband beam may cover other users at different subcarriers. Accordingly, the analog beam-focusing vector at the central frequency $f_c^k$ is expressed as
\begin{equation}
\label{eq_beam_focusing}
\begin{aligned}
&\mathbf w\left(f_c^k,\mathbf p_i,\{\mathbf p_t^n\}_{n=1}^{N_t}\right)\\
& = \left[e^{-j\frac{2\pi f_c^k}{c}|\mathbf p_i-\mathbf p_t^1|_2},\ldots,e^{-j\frac{2\pi f_c^k}{c}|\mathbf p_i-\mathbf p_t^{N_t}|_2}\right].
\end{aligned}
\end{equation}
To identify the specific users covered by each subcarrier, it is essential to characterize the spatial distribution of the wideband beam. Ideally, the drifted focus position $(r_m^k, \theta_m^k,\phi_m^k)$ for each subcarrier $f_m^k$ would be derived analytically. However, for a near-field wideband beam generated by a two-dimensional planar aperture, this derivation is generally intractable because the two aperture dimensions generally do not share one common drifted focal range. This phenomenon is stated in the following theorem.

\textbf{Theorem 1:} Consider a general two-dimensional planar array, such as a standard UPA, a sparse planar array, or a planar array-of-subarrays, whose antenna elements are distributed along two aperture dimensions. For a wideband near-field beam generated by this array, a subcarrier away from the central frequency generally produces different candidate focal ranges along the two aperture dimensions. This dimension-dependent focal mismatch, referred to as the astigmatism phenomenon, prevents the beam from being perfectly refocused at a single spatial point. \emph{Proof: See Appendix.}

Theorem 1 complements the existing results obtained for ULA and UCA geometries, as introduced in Sec.~\ref{section_intro}, by extending the analysis to a general two-dimensional planar aperture~\cite{rainbow,UCA}. It is worth noting that Theorem 1 does not imply that the gain map has no maximum point. A numerical peak always exists for a finite array and a finite observation region. However, this gain peak is not generally associated with a unique perfectly phase-matched focal point for two-dimensional arrays. At an off-center subcarrier, adjusting the observation point may compensate for the frequency-dependent phase variation along one aperture dimension, while generally leaving residual phase variation along the other. Consequently, the contributions from all antenna elements cannot be fully coherently combined at one common spatial point. Therefore, for user grouping, we do not first calculate a drifted focal point and then infer the coverage region around it. Instead, we directly evaluate the actual beamforming gain at each candidate user position and define the effective beam split coverage by the 3-dB gain condition. Specifically, at an arbitrary user position $\mathbf{p}_u$ with subcarrier $f_m^k$, the beamforming gain is calculated as
\begin{equation}
\label{eq:Gain}
\begin{aligned}
    & \text{Gain}(\mathbf{g}_u,\mathbf{p}_u,f_m^k) \\
    = & \left|\frac{1}{N_t}\mathbf{w}^H(f_c^k,\mathbf{g}_u,\{\mathbf{p}_n^t\})\mathbf{a}(f_m^k, \mathbf{p}_u,\{\mathbf{p}_n^t\})\right|^2\\
    = & \left|\frac{1}{N_t}\sum_{n=1}^{N_t}e^{j\frac{2\pi}{c}\left[f_{c}^k\rVert \mathbf{g}_{u}-\mathbf{p}_t^n\rVert_2-f_m^k\lVert \mathbf{p}_u-\mathbf{p}_t^n\rVert_2\right]}\right|^2.
\end{aligned}
\end{equation}
We define the beam split coverage of a wideband beam as the union of the 3-dB regions produced by all its subcarriers. Specifically, for a wideband beam targeting $\mathbf{g}_u$ in sub-band $k$, its coverage region is defined as
\begin{equation}
\label{eq:split_coverage}
    \mathcal{C}^k(\mathbf{g}_u) = \{\mathbf{p} : \exists m \in \{1,...,M\}, \text{Gain}(\mathbf{g}_u,\mathbf{p},f_m^k) \geq 1/2\}.
\end{equation}%
To provide a numerical illustration of this definition, Fig.~\ref{fig:3dB_coverage} shows the beamforming gain of a 1024-antenna UPA. The array lies on the x-z plane and is centered at the origin, while the target is located at $\mathbf p_c=(0.05,0.20,0)$~m, which lies in the near-field region of the considered array. The analog beamforming vector is designed at the central frequency of 300 GHz and the beamforming gain is calculated via \eqref{eq:Gain} at 270, 300, and 330 GHz frequency, respectively. The relatively large frequency separation is adopted only to make the frequency-dependent variation of the beam pattern clearly visible. In this figure, the white curves indicate the 3-dB contours at the corresponding frequencies. The circle marks the intended target position $\mathbf p_c$, whereas the cross marks the position of the numerically obtained maximum-gain. At the central frequency, the maximum-gain position coincides with the intended target, while at a non-central frequency, the frequency-dependent phase mismatch shifts the peak position and the 3-dB region. Moreover, the 3-dB regions produced by different subcarriers may partially overlap, and their union forms the beam-split coverage of the entire wideband beam. Accordingly, user $u$ can be assigned to a group with center $\mathbf{g}_u$ if $\mathbf{p}_u$ in $\mathcal{C}^k(\mathbf{g}_u)$.

To reduce the power consumption, we propose to minimize the required number of RF chains. Therefore, the user grouping problem is expressed as
\begin{subequations}
\label{eq:problem_user_grouping}
\begin{align}
\mathcal{P}_2: \underset{\{\mathbf{g}^u\}} {\operatorname{minimize}} &\ \ \ |\{\mathbf{g}_u\}_{u=1}^U|, \\
    \text{s.t.} \quad & \mathbf{p}_u \in \mathcal{C}^k(\mathbf{g}_u), \forall u\in\mathcal{U}_k\\
    & \mathbf{g}_u \in \{\mathbf{p}_i:i\in\mathcal{U}_k\}
    \label{eq_constr_group_center}
\end{align}
\end{subequations}
where $\{\mathbf{g}_u\}_{u=1}^U$ denotes the set of group centers, and $|\{\mathbf{g}_u\}_{u=1}^U|$ represents the number of groups, which equals $L_t$. Constraint~\eqref{eq_constr_group_center} indicates that the group center should be in the same sub-band as the users.
Next, we show that $\mathcal{P}2$ is equivalent to a directed minimum dominating set (MDS) problem. For each sub-band $k$, we construct a directed graph $\mathcal{G}_k=(V_k,E_k)$, where $V_k=\mathcal{U}_k$ is the set of users assigned to sub-band $k$, and
\begin{equation}
    E_k = \{(i,j)\in V_k\times V_k|\exists\ m, \ \operatorname{Gain}\left(\mathbf{p}_i,\mathbf{p}_j, f_m^k\right)\geq \frac{1}{2}\}.
\end{equation}
Therefore, an edge \((i,j)\in E_k\) indicates that user \(j\) can be assigned to a group with group center \(\mathbf{p}_i\). Hence, a set of selected group anchors \(D_k\subseteq V_k\) yields a feasible grouping if and only if every vertex \(j\in V_k\) either belongs to \(D_k\) or is in the outgoing neighborhood of at least one vertex in \(D_k\). This is precisely the definition of a dominating set of \(\mathcal{G}_k\). Since each selected group center corresponds to one user group, minimizing the number of groups in \(\mathcal{P}2\) is equivalent to finding a minimum dominating set in each \(\mathcal{G}_k\). Therefore, the problem can be efficiently solved by a greedy algorithm. The algorithm is summarized as follows:
\begin{itemize}
\item Step 1: For each user pair $u_i$ and $u_j$ within sub-band $k$, calculate $\text{Gain}(\mathbf{p}_{u_i},\mathbf{p}_{u_j},f_m^k)$ for each subcarrier $f_m^k$. A directed edge from vertex $u_i$ to $u_j$ is established if constraint (\ref{eq:split_coverage}) is satisfied, forming the directed graph $(V_k,E_k)$.
\item Step 2: Iteratively select the ungrouped user vertex with the highest out-degree (i.e., with the most outgoing edges) as a group center. All users pointed to by this center are assigned to its group. 
\item Step 3: Any remaining ungrouped users with no connecting edges are assigned to their own individual groups.
\end{itemize}
The proposed user-grouping strategy is directly based on the beamforming gain and is therefore not limited to the LoS model. For a multipath channel, the coverage graph can be constructed in the same manner by recalculating the gain in \eqref{eq:Gain} using the multipath channel vector and a multipath-aware beamforming vector.

Then, with the grouped users, the subcarriers can be allocated to maximize the minimum sum beamforming gain of each user. Assuming perfect interference cancellation at the digital stage, we have
\begin{subequations}
\label{eq:problem_split_multiplexing}
\begin{align}
    \max_{\mathcal{M}_u} &\ \min\sum_{\substack{m \in \mathcal{M}_{u}, \\ u \in \mathcal{U}_g}} \text{Gain}(\mathbf{g}_u,\mathbf{p}_u,f_m^k), \\
    \text{s.t.}\ \ &\mathcal{M}_u \neq \varnothing, \forall u \in \mathcal{U}_g, \label{eq:non_empty_subcarrier}\\
    & \mathcal{M}_i \cap \mathcal{M}_j = \varnothing,\forall i,j \in \mathcal{U}_{g},
\end{align}
\end{subequations}
where $\mathcal{U}_{g}$ denotes the set of users in group $g$. According to eq.~(\ref{eq:non_empty_subcarrier}), each user should be assigned at least one subcarrier. Therefore, an obvious strategy is to assign the central frequency $f_c^k$ to each group center user, and other users within the group are assigned the subcarrier that yields the highest $\text{Gain}$. Subsequently, an iterative allocation process is conducted where in each iteration the user with the lowest current gain is identified. Then, given the limited number of available subcarriers, the exhaustive search can be performed to assign the best available subcarrier to this user. The overall two-phase frequency allocation algorithm is summarized in Algorithm~\ref{alg:two-phase}.

The computational complexity of the proposed two-phase frequency-allocation algorithm is analyzed as follows. In Phase I, calculating the pairwise channel correlations, sorting them, and scanning the sorted correlation list require a complexity of $O(N_tU^2)$, $O(U^2\log U)$, and $O(KU^2)$, respectively. As $\log U$ and $K$ are generally smaller than $N_t$, the total complexity of Phase I is $O(N_tU^2)$. In Phase II, constructing the gain-based coverage graphs has a complexity of $O(MN_t\sum_{k=1}^{K}|\mathcal{U}_k|^2)$. The greedy solution to the directed MDS problem has complexity $O(\sum_{k=1}^{K}(U_k+|E_k|)\log U_k)$, where \(|E_k|\) denotes the number of directed edges in sub-band $k$. Therefore, the overall complexity of Algorithm~\ref{alg:two-phase} is $O(N_tU^2+MN_t\sum_{k=1}^{K}U_k^2+\sum_{k=1}^{K}(U_k+|E_k|)\log U_k)$.

\begin{algorithm}
	\renewcommand{\algorithmicrequire}{\textbf{Input:}}
	\renewcommand{\algorithmicensure}{\textbf{Output:}}
	\caption{Two-phase Frequency Allocation}
	\label{alg:two-phase}
	\begin{algorithmic}[1]
		\REQUIRE Frequency set $\mathcal{F} = \{f_m^k\}$, user positions \{$\mathbf{p}_u\}$.
	    \STATE Solve problem $\mathcal{P}_1$ and obtain $\mathcal{U}_k, \forall k$.
        \STATE Solve problem $\mathcal{P}_2$ and obtain $\{\mathbf{g}_u\}_{u=1}^G$.
        \STATE Calculate $\text{Gain}(\mathbf{g}_u,\mathbf{p}_u,f_m^k)$ via \eqref{eq:Gain} for each group.
        \STATE Allocate subcarrier with the highest gain to the target user, and remove the allocated subcarriers from $\mathcal{F}$
        \FOR{$g = 1$ to $G$}
        \REPEAT
        \STATE For $u \in \mathcal{U}_g$, identify the user with the lowest gain, and add the best subcarrier $f_m^k$ to $\mathcal{M}_u$.
        \STATE Remove $f_m^k$ from $\mathcal{F}$.
        \STATE Recalculate $\sum_{m \in \mathcal{M}_u}\text{Gain}(\mathbf{g}_u,\mathbf{p}_u,f_m^k)$.
        \UNTIL All subcarriers are allocated.
        \ENDFOR
		\ENSURE $\mathcal{M}_u$.
	\end{algorithmic} 
\end{algorithm}

\section{Dynamic Antenna Selection and MA Position Optimization}
\label{sec_problem_P3}
With the inter-user interference largely mitigated by the proposed frequency allocation strategy, we now aim to maximize the sum rate through spatial domain multiplexing. Specifically, dynamic antenna selection is first employed to establish the active array topology. Based on this configured architecture, we then jointly optimize the MA positions and hybrid beamforming weights to optimize the array geometry and fully exploit the spatial multiplexing gain.

\subsection{Problem Formulation}
Based on the established signal model and design objectives, the optimization problem is formulated as
\begin{subequations}
        \begin{align}
            \underset{\{\mathbf{p}_{\rm sub}^n\}, \{\mathbf{p}_{u}\}, \mathbf{S}, \mathbf{F}^k_{\mathrm{RF}}, \mathbf{F}_{\mathrm{BB}}}{\operatorname{maximize}} & \ R = \sum_{k=1}^K\sum_{m = 1}^M\sum_{u \in \mathcal{U}^{k}_m}B_m^k SE_u^k[m] \label{R in Original}\\
            \text { s.t. } \quad \quad
            & C_1: \lVert\widetilde{\mathbf{F}}^k_{\rm RF}\mathbf{F}^k_{\rm BB}[m]\rVert^2_{\rm F} = |\mathcal{U}_m^k|,\\
            & C_2: \mathbf{p}^n_{\rm sub} \in \mathcal{A}_{sub}^n, \forall n,\\
            & C_3: \mathbf{p}_{u} \in \mathcal{A}_{u}, \forall u,\\
            & C_4: \mathbf{S}[i,j] \in \{0,1\}, \forall i,j, \\
            & C_5: \left|\mathbf{F}_{\mathrm{RF}}[i,j]\right|^{2}=1, \forall i, j, \label{eq:FRF_constraint_FC}
        \end{align}
    \label{Original problem}%
\end{subequations}
where $B_m^k$ is the bandwidth of each subcarrier, which is set as $B_m^k = \frac{B}{M}$ $\forall m$ in this work. Constraint $C_1$ is the normalized power constraint for the hybrid precoders where $|\mathcal{U}_m^k|$ represents the number of users served by $f_m^k$. $C_2$ and $C_3$ are the movable region constraints for the transmitter and the user, respectively, where $\mathbf{p}_{\rm sub}$ and $\mathbf{p}_{u}$ denote the position of the reference antenna in each movable subarray and each user. $\mathcal{A}_{\rm sub}^n$ and $\mathcal{A}_{u}$ denote the feasible regions of the $n^{\rm th}$ movable subarray and each user $u$, respectively. $C_4$ and $C_5$ denote the hardware constraints for the switch network matrix and the analog precoder implemented by phase shifters. 
To solve this non-convex optimization problem, we decompose the original problem into two sub-problems. We first consider the design of the dynamic antenna selection, and then consider the joint optimization of the antenna position and hybrid beamformers.

\subsection{Dynamic Antenna Selection}
\label{sec_antenna_selection}
\begin{figure}
    \centering
    \includegraphics[width=0.43\textwidth]{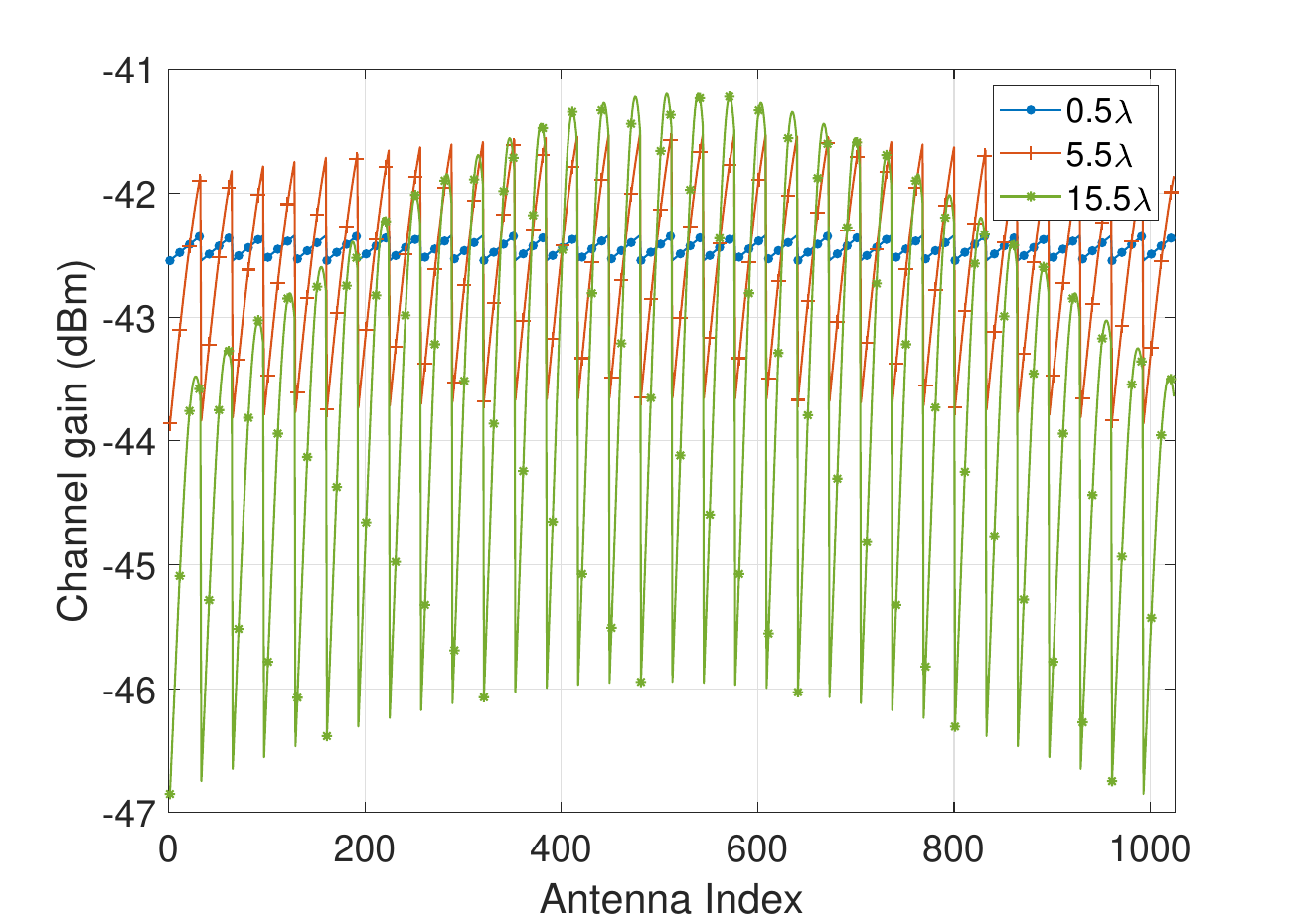}
    \caption{Channel gain versus antenna index.}
    \label{Fig_antenna_contribution}
\end{figure}

The expansion of antenna array dimensions in massive MIMO systems introduces significant spatial variations in per-antenna channel gains to individual users. As shown in Fig.~\ref{Fig_antenna_contribution}, the disparity in channel gain between the highest-contribution antenna and the lowest-contribution antenna increases with antenna spacing. For half-wavelength spacing, all antennas exhibit nearly identical gains. However, when the spacing expands to $5.5\lambda$, the peak-to-minimum gain difference rises to  around $2$~dBm, and for 15.5$\lambda$ spacing, the gap exceeds $5$~dBm. This phenomenon implies that distant antennas contribute negligible signal energy while consuming full power. Consequently, proactively selecting high-gain antennas can effectively improve the energy efficiency. 
Leveraging this spatial disparity, we propose a greedy antenna-selection method. For user group $g$ in sub-band $k$, let $h_{g,n}^{k}$ denote the channel coefficient between the $n^{\rm th}$ transmit antenna and the group center, and let $\mathcal{A}_{g}^{k}$ denote the set of selected antennas. Then, the fraction of the aggregate channel contribution retained after antenna selection is defined as
\begin{equation}
\label{eq:zeta}
    \zeta_{g}^{k} =\frac{\sum_{n\in\mathcal{A}_{g}^{k}}\left|h_{g,n}^{k}\right|}
{\sum_{n=1}^{N_t}\left|h_{g,n}^{k}\right|}.
\end{equation}
Starting from the antennas closest to the group center, antennas are progressively activated until an antenna selection threshold is achieved as $\zeta_{g}^{k}\geq\gamma_{\mathrm{th}}$. Therefore, $\gamma_{\mathrm{th}}$ specifies the fraction of the total channel contribution to be retained after antenna selection. A larger $\gamma_{\mathrm{th}}$ activates more antenna connections, whereas a smaller value disconnects more weak-contribution antenna branches. As this method sorts the distances of the $N_t$ antennas for each of the $G$ group centers, the overall complexity is $O(GN_t\log N_t)$.

\subsection{Joint Precoding and MA Optimization}
In this section, the hybrid precoder and the positions of movable antennas are jointly optimized through the PSO algorithm to maximize the sum rate. The algorithm involves two key steps: 1) optimization of the hybrid precoder given fixed antenna positions, and 2) optimization of the antenna positions  while keeping the precoder fixed.

\subsubsection{Hybrid Precoding}
As users within the same group are served by a common wide beam generated from one RF chain, we propose to design the analog precoder with the objective of focusing the beam towards the group center. Therefore, by adopting the beam focusing vector, the analog precoder $\mathbf{F}_{\rm RF}^k$ across all $M$ subcarriers can be represented as
\begin{equation}
\label{eq:FRF}
\mathbf{F}_{\rm RF}^k = [\mathbf{w}(f_c^k,\mathbf{g}_1,\{\mathbf{p}_t^n\}_{n=1}^{N_t}), \cdots, \mathbf{w}(f_c^k,\mathbf{g}_G,\{\mathbf{p}_t^n\}_{n=1}^{N_t})].
\end{equation}
Then, for users at subcarrier $f_m^k$, we apply the block diagonalization (BD) method at the baseband to obtain the corresponding $\mathbf{f}_{\rm BB,u}^k[m]$, which mitigates all the residual user interference~\cite{HBD,HBF}. Specifically, the digital precoder for the $u^\mathrm{th}$ user lies in the null space of other users' channel matrix, i.e., $\hat{\mathbf{h}}_u^k[m]\mathbf{f}_{\mathrm{BB,i}}^k[m] \approx 0, \forall i \neq u$, where $\hat{\mathbf{h}}_u^k[m] = \mathbf{h}_u^k[m] \mathbf{F}^k_{\rm RF}$. After obtaining the BD-based digital beamforming direction, each hybrid beamforming vector is normalized to unit power, and the power $p_u^k[m]$ is subsequently applied according to the equal power allocation strategy defined in Sec.~\ref{section_Sys_channel_Model_wideband}.

\subsubsection{PSO-based MA Optimization}
\label{sec_MA_optimization}

\begin{figure}
    \centering
    \includegraphics[width=0.43\textwidth]{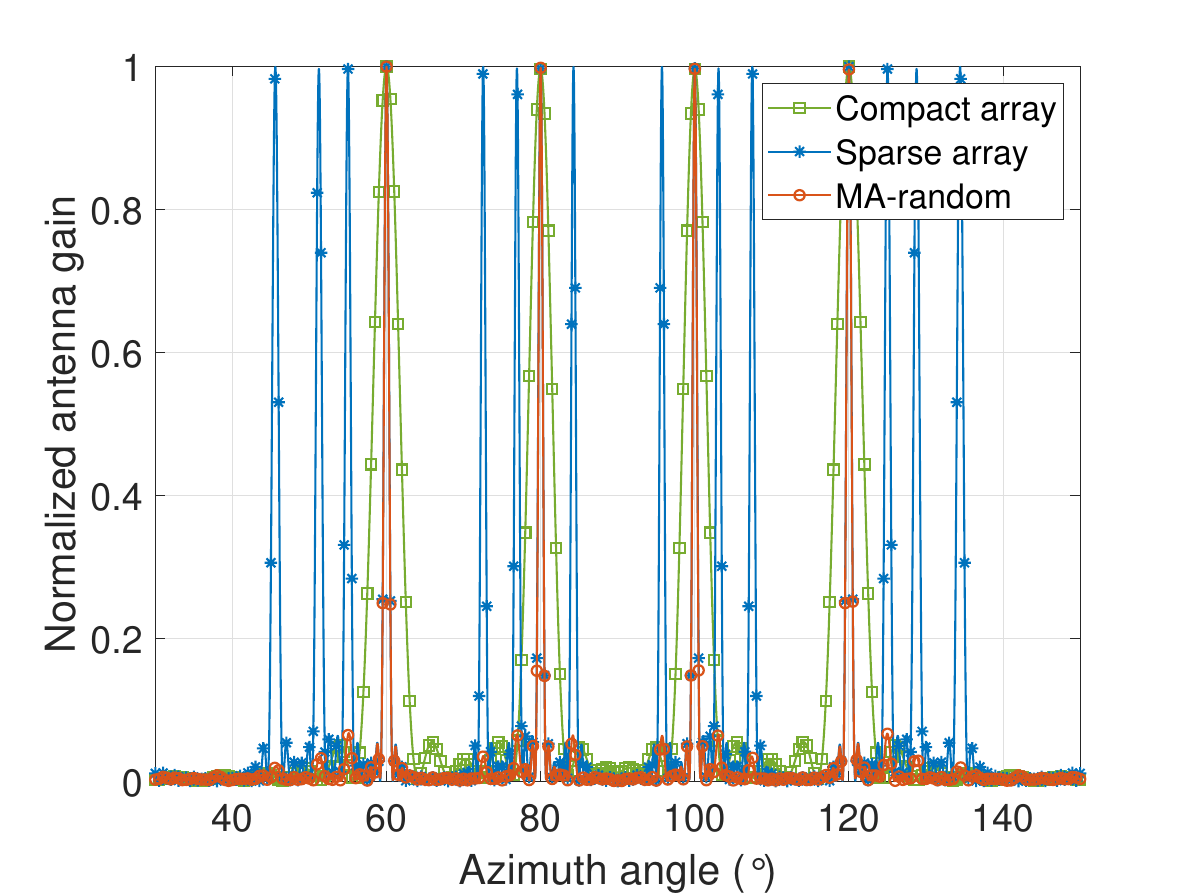}
    \caption{Channel gain for different antenna configurations.}
    \label{fig_grating_lobe}
\end{figure}

Optimizing the MA position brings two advantages for multi-user communications. On one hand, the MA positions influence the channel matrix directly, which in turn affects the overall system performance. On the other hand, conventional sparse arrays with uniform antenna spacing inherently suffer from grating lobes due to the periodicity of the array factor (AF). Fig.~\ref{fig_grating_lobe} compares the beam pattern for different antenna array configurations where target users are located at $60^{\circ},80^{\circ},100^{\circ},120^{\circ}$. For a traditional compact array with half-wavelength antenna spacing, four beams are generated towards each user with the largest beamwidth. For a sparse array where adjacent elements are spaced at intervals exceeding $\lambda/2$, the AF exhibits periodic maxima where constructive interference occurs at unintended positions. This spatial aliasing causes multiple high-gain sidelobes, degrading beamforming precision and increasing interference. Fortunately, through non-uniform position optimization of the movable subarrays, this grating lobe effect can be mitigated. By displacing antennas to break spatial periodicity, the phase relationships across the array become aperiodic such that electromagnetic waves no longer superimpose constructively at periodic grating-lobe angles. 
To harness these advantages, we optimize the MA positions to maximize the sum rate under constraint $C_2$ and $C_3$.

Given the hybrid beamformers, the interference can be neglected such that the sum rate in \eqref{R in Original} can be simplified as
\begin{equation}
\begin{aligned}
    R = \sum_{k=1}^K\sum_{m = 1}^MB_m^k\log _{2} & \operatorname{det}\left( \mathbf{I}+\sigma_{\mathrm{n}}^{-2}\mathbf{H}^k[m]\mathbf{F}_{\rm RF}^k\mathbf{F}_{\rm BB}^k[m]\right.\\
    & \left. \times (\mathbf{F}_{\rm BB}^k[m])^H(\mathbf{F}_{\rm RF}^k)^H (\mathbf{H}^k[m])^H\right),
\end{aligned}
\label{eq_simplified_R}
\end{equation}
where $\mathbf{H}^k[m]$ and $\mathbf{F}_{\rm BB}^k[m]$ denote the channel matrix and the digital precoder for all users served by subcarrier $f_m^k$, i.e.,
\begin{subequations}
\begin{align}
    & \mathbf{H}^k[m] = \left[(\mathbf{h}_1^k[m])^T,\cdots, (\mathbf{h}_{U}^k[m])^T\right]^T, u \in \mathcal{U}_m^k,\\
    & \mathbf{F}_{\rm BB}^k[m] = \left[\mathbf{f}_{\rm BB,1}^k[m],\cdots, \mathbf{f}_{\mathrm{BB},U}^k[m]\right],u \in \mathcal{U}_m^k.
\end{align}
\end{subequations}
To solve this high-dimensional and non-convex optimization problem which is prone to local optima, the PSO algorithm is employed for its robust global search capability. 
Specifically, each particle represents a potential configuration of the positions of the reference antennas for each subarray at the BS, and the receiver antennas in the 3-dimensional space, i.e., each particle consists of $3(N_{\rm sub}+U)$ coordinates. The initial position of the $q^{\rm th}$ particle is set as
\begin{equation}
    \mathbf{p_q^{(0)}} = [(\mathbf{p}_t^1)^0,\cdots,(\mathbf{p}_t^{N_{\rm sub}})^0, \mathbf{p}_1, \cdots, \mathbf{p}_U] + \mathscr{U}(\mathbf{lb},\mathbf{ub}),
\end{equation}
where $\mathbf{lb}$ and $\mathbf{ub}$ denote the vectors for the lower and the upper bound of the movable region for each antenna at the transceiver, and $\mathscr{U}$ denotes the uniform distribution. The velocity update for the $t_{\rm th}$ iteration is represented as
\begin{equation}
\label{eq:velocity_update}
    \mathbf{v}_q^{(t+1)} = \omega\mathbf{v}_q^{(t)}+c_1r_1(\mathbf{p}_{q}^{local}-\mathbf{p}_q^{(t)})+c_2r_2(\mathbf{p}_{q}^{global}-\mathbf{p}_q^{(t)}),
\end{equation}
where $r_1$ and $r_2$ are random factors in $[0,1]$ that help the particles escape local optima. $c_1$ and $c_2$ are the individual and global learning rates, respectively, and $\omega$ denotes the inertia weight which balances global exploration and local exploitation capabilities. $\mathbf{p}_q^{local}$ and $\mathbf{p}^{global}$ denote the best position of the $q^{\rm th}$ particle and the best position of all particles, respectively.
Then, the particle update is represented as
\begin{equation}
\label{eq:particle_update}
    \mathbf{p_q^{(t+1)}} = \mathbf{p_q^{(t)}}+\mathbf{v}_q^{(t+1)}.
\end{equation}
The algorithm of antenna selection and the joint optimization is summarized in Algorithm~\ref{Ag: beamforming_MA}.
\begin{algorithm}
	\renewcommand{\algorithmicrequire}{\textbf{Input:}}
	\renewcommand{\algorithmicensure}{\textbf{Output:}}
	\caption{Antenna selection and MA position optimization}
	\label{alg2}
	\begin{algorithmic}[1]
		\REQUIRE Antenna selection threshold $\gamma_{th}$ and positions of group centers.
        \STATE Sort the antennas based on the distance between the antenna and the group center for each group $g$.
        \FOR{$g=1$ to $G$}
        \REPEAT
        \STATE Select the first un-selected antenna for group $g$ and update $\zeta_g^k$ via \eqref{eq:zeta}.
        \UNTIL $\zeta_g^k \geq \gamma_{\rm th}$
        \ENDFOR
        \STATE Initialize antenna position particles $\{\mathbf{p}_n^{(0)}\}$, velocities $\mathbf{v}_n^{(0)}$.
        \REPEAT
        \FOR{$n = 1$ to $N_{par}$}
		\STATE Construct analog beamforming matrix $\mathbf{F}_{\rm RF}^k$ through \eqref{eq:FRF}.
		\STATE Calculate $\mathbf{F}_{\mathrm{BB}}$ through the BD algorithm.
        \STATE Update the fitness function \eqref{eq_simplified_R}.
        \STATE Update velocity and particle position via \eqref{eq:velocity_update} and \eqref{eq:particle_update}.
        \ENDFOR
        \UNTIL The stopping criterion is triggered.
		\ENSURE $\mathbf{F}_{\rm BB}$, $\mathbf{F}_{\rm RF}$,$\{\mathbf{p}_t^n\}_{n=1}^{N_t}$,$\mathbf{p}_u$
	\end{algorithmic} 
 \label{Ag: beamforming_MA}
\end{algorithm}

\begin{table}
\centering
\caption{Simulation Setup.}
\label{Table: Simulation}
\begin{tabularx}{\linewidth}{|>{\centering\arraybackslash}p{5cm}|>{\centering\arraybackslash}X|}
\hline
\textbf{Parameter}& \textbf{Value}\\
\hline
Total bandwidth B & $30$~GHz\\
Available central frequencies $f_c^k$ & $300, 305, 310$, $315,320,325$~GHz\\
Sub-band bandwidth & $5$~GHz\\
Subcarrier bandwidth & $1$~GHz\\
Number of multipath components $N_p$ & 1\\
Number of transmit antennas $N_t$ & 1024\\
Number of receive antennas $N_r$ & 1\\
Number of data streams for each user $N_s$ & 1\\
Number of users $U$ & 50\\
Number of movable subarrays $N_{\rm sub}$ & 16\\
Antenna movable region $p\lambda$ & $p$ = 10\\
Maximum number of iterations for PSO-based algorithm & $50$\\
$c_1$ and $c_2$ for PSO-based algorithm & $1.7,1.7$\\
Swarm size for PSO-based algorithm & 40\\
Antenna selection threshold & 0.8\\
Monte Carlo trials & $1000$\\
\hline
\end{tabularx}
\end{table}

\section{Performance Evaluation}
\label{section_performance}
In this section, we first analyze the parameters for the proposed algorithms and then we compare the sum rate and EE performance of the proposed D-FPFA schemes with different benchmarks. The simulation setup is summarized in Table~\ref{Table: Simulation}.

\subsection{Parameter Evaluation}
\label{sec_parameter_eval}

\begin{figure*}
\centering
\begin{minipage}{0.32\linewidth}
\centering
\includegraphics[width=\linewidth,height = 0.75\linewidth]{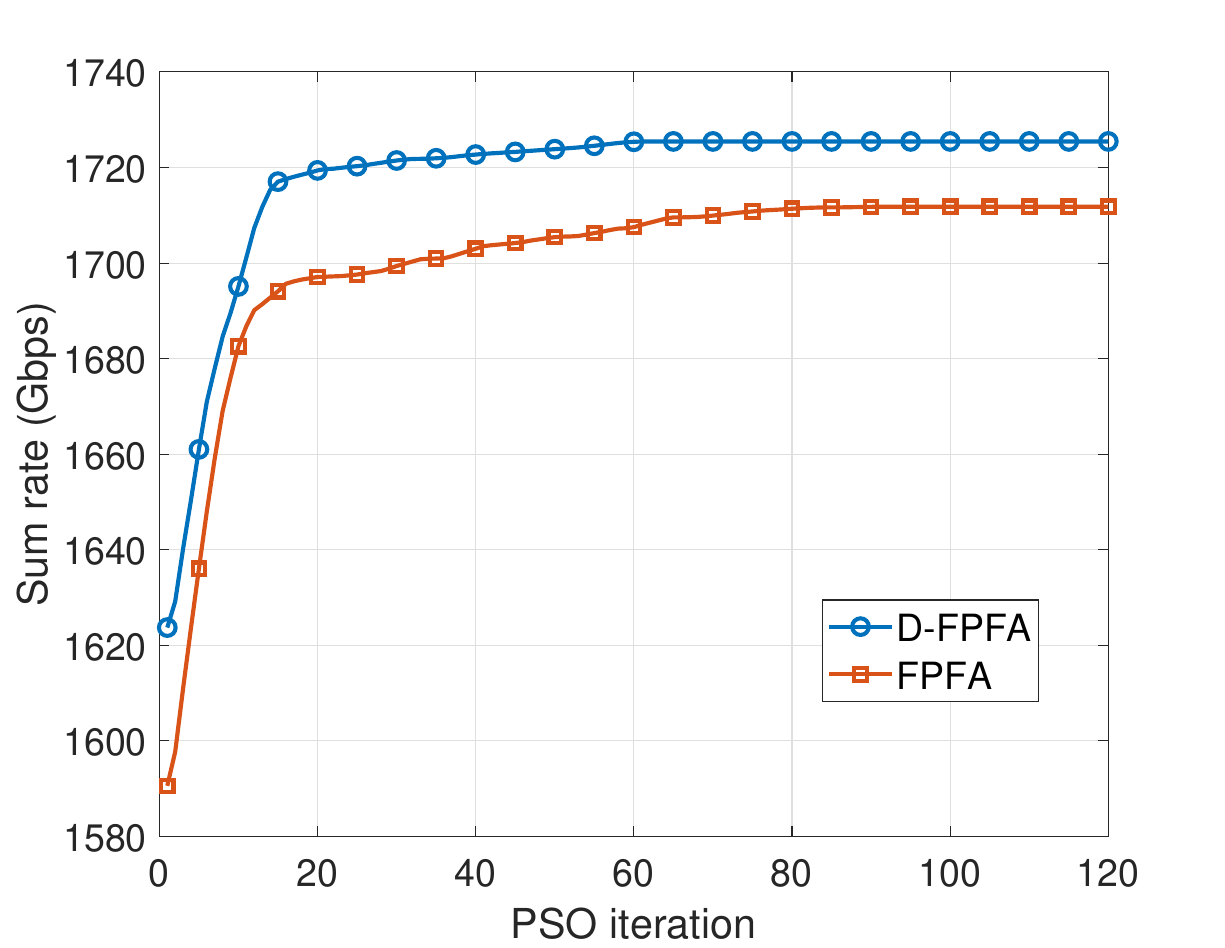}
\captionof{figure}{Convergence behavior of PSO-based position optimization.}
\label{fig:convergence}
\end{minipage}\hfill
\begin{minipage}{0.32\linewidth}
\centering
\includegraphics[width=\linewidth,height = 0.75\linewidth]{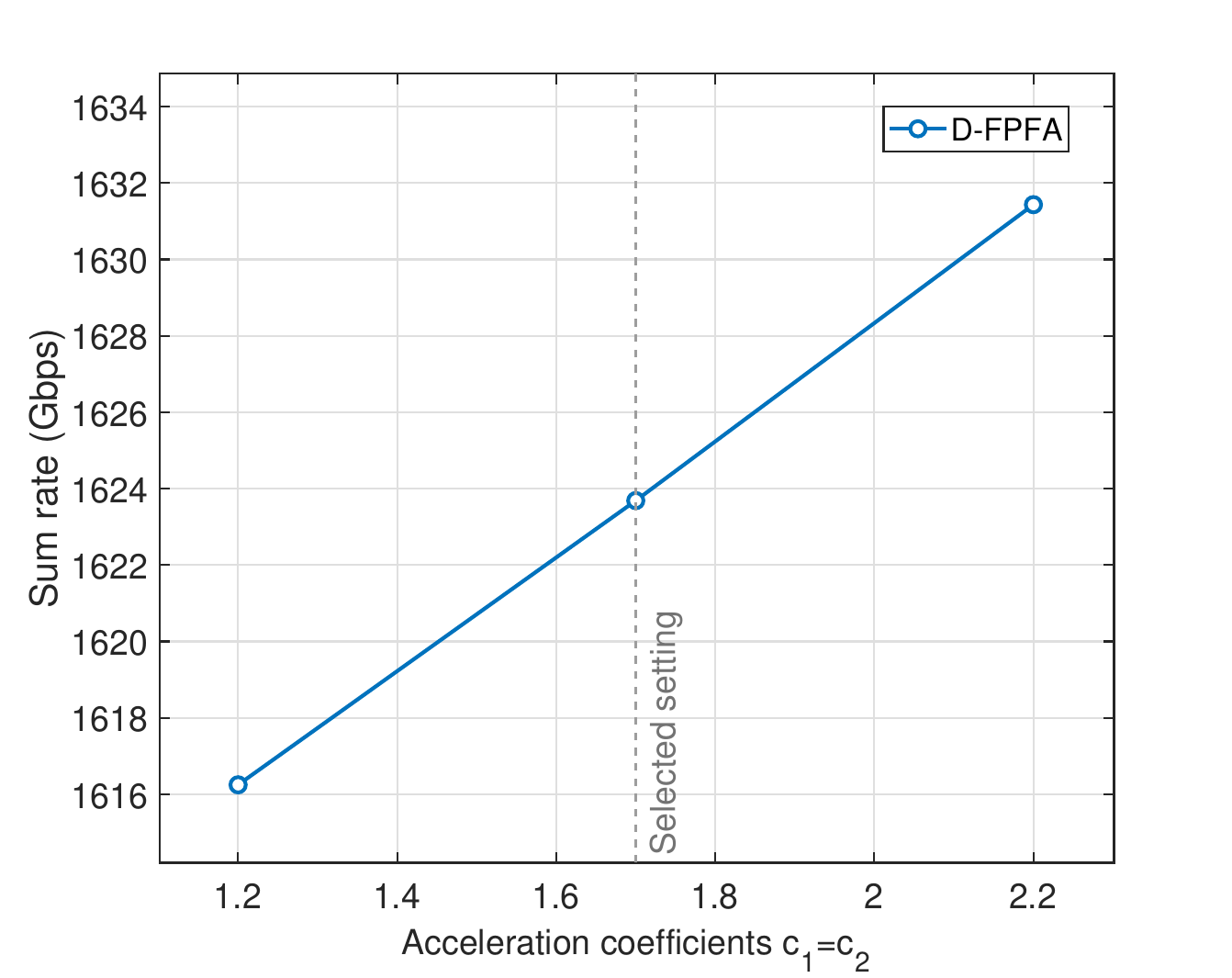}
\captionof{figure}{Sensitivity analysis of acceleration coefficients.}
\label{fig:c1c2}
\end{minipage}
\begin{minipage}{0.32\linewidth}
\centering
\includegraphics[width=\linewidth,height = 0.75\linewidth]{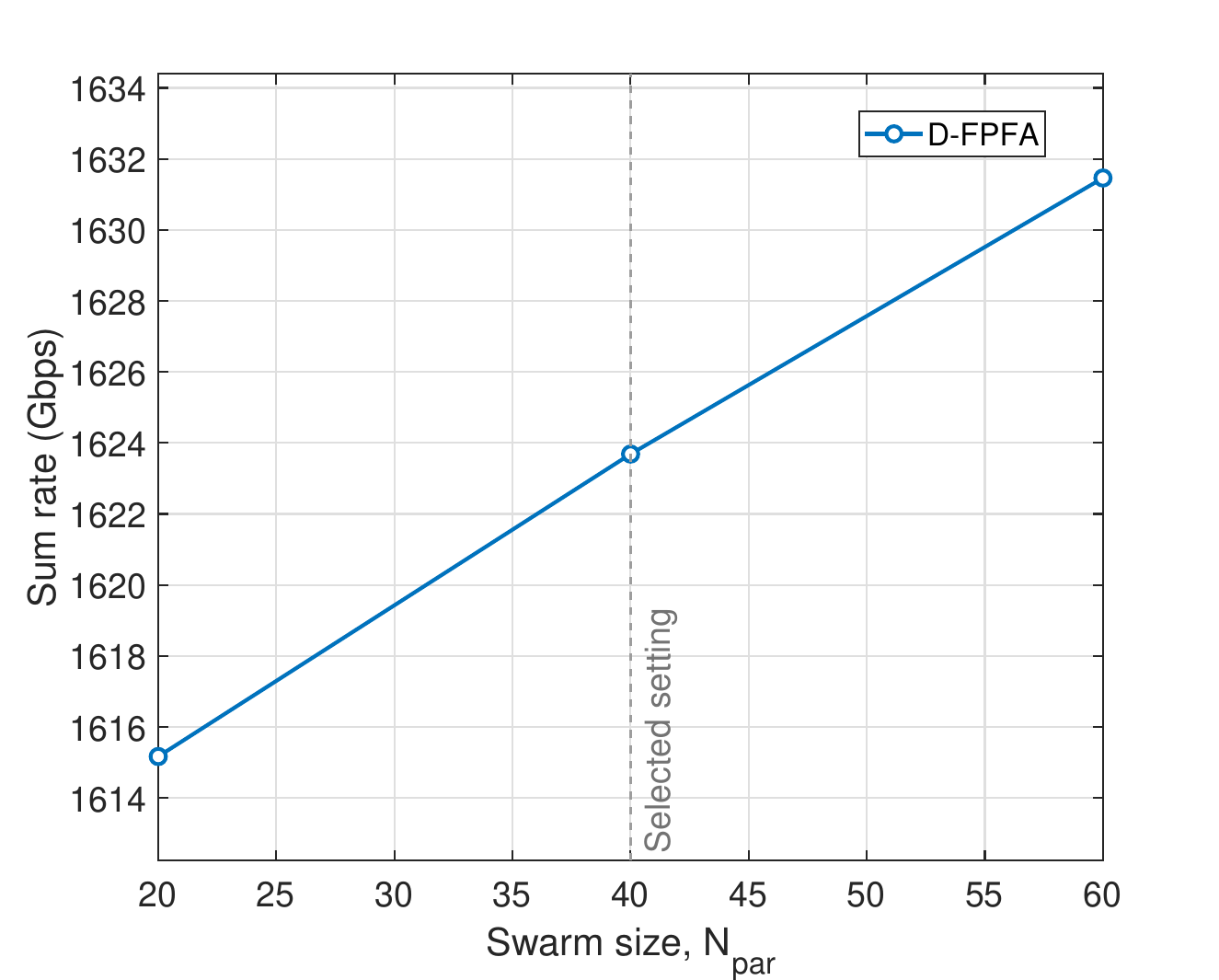}
\captionof{figure}{Sensitivity analysis of the swarm size.}
\label{fig:swarmsize}
\end{minipage}
\end{figure*}

\begin{figure*}
\centering
\begin{minipage}{0.32\linewidth}
\centering
\includegraphics[width=\linewidth,height = 0.75\linewidth]{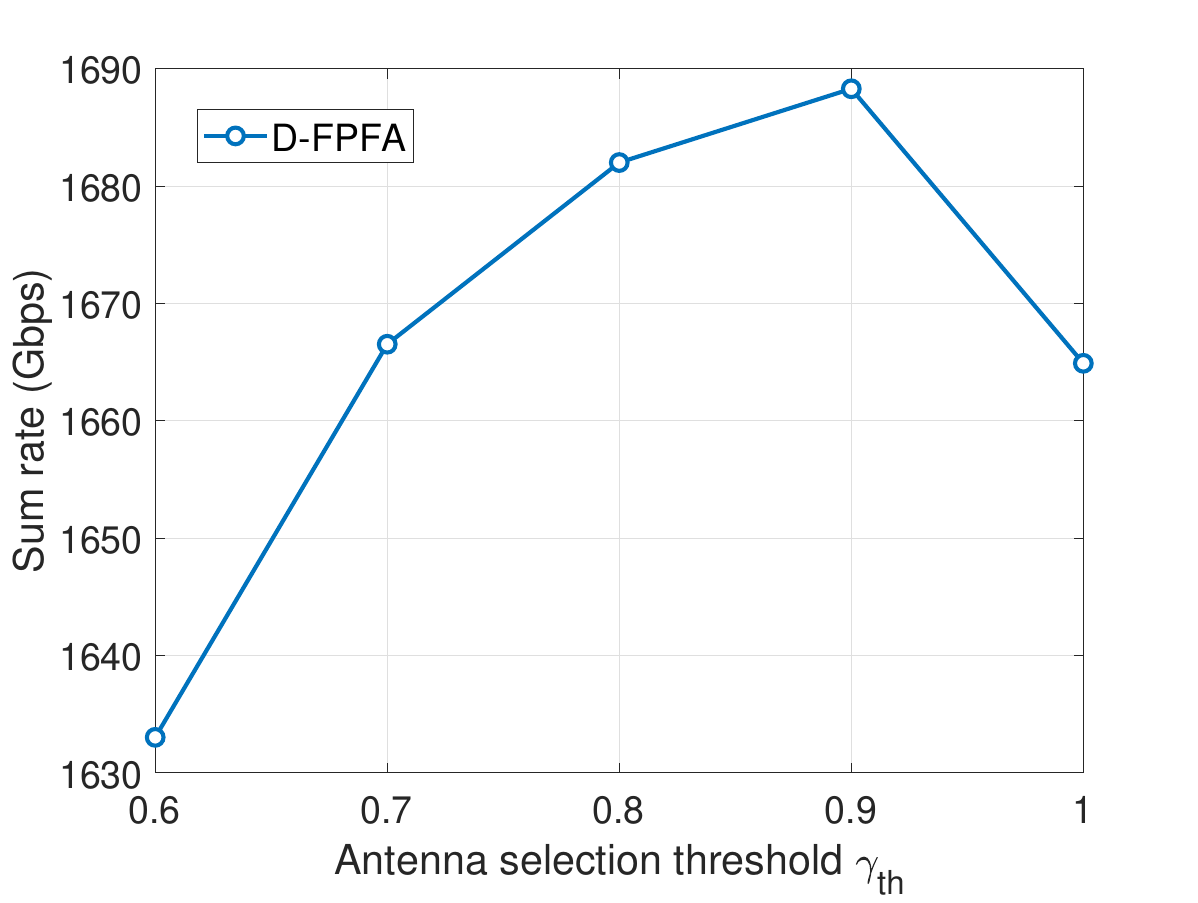}
\captionof{figure}{Sum rate versus the antenna selection threshold.}
\label{fig:SE_vs_gamma}
\end{minipage}\hfill
\begin{minipage}{0.32\linewidth}
\centering
\includegraphics[width=\linewidth,height = 0.75\linewidth]{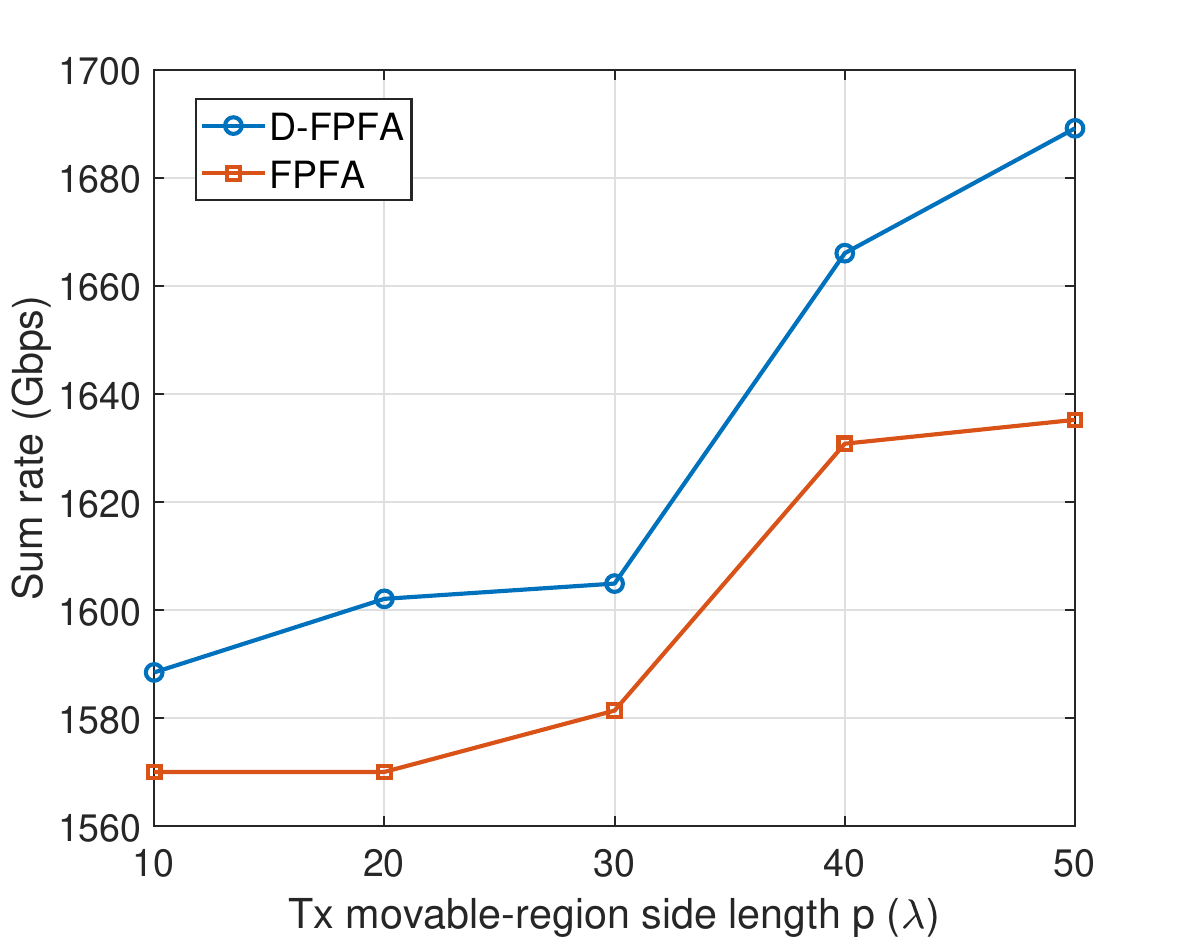}
\captionof{figure}{Sum rate versus the size of movable region.}
\label{fig:SE_vs_movableregion}
\end{minipage}
\begin{minipage}{0.32\linewidth}
\centering
\includegraphics[width=\linewidth,height = 0.75\linewidth]{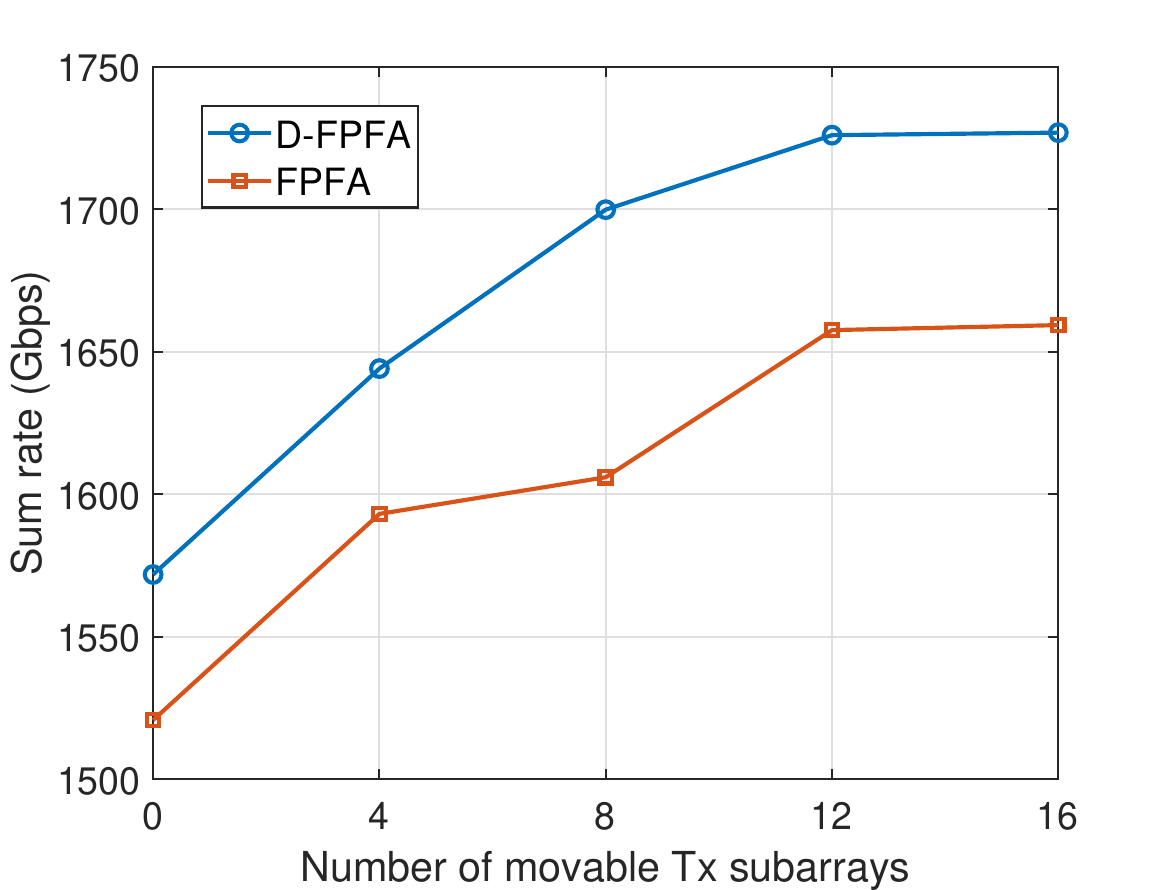}
\captionof{figure}{Sum rate versus the number of movable subarrays.}
\label{fig:SE_vs_numMA}
\end{minipage}
\end{figure*}

We first evaluate the convergence behavior of the PSO-based movable-array position optimization. As shown in Fig.~\ref{fig:convergence}, the sum rate of both D-FPFA and FPFA increase rapidly during the initial iterations, with most of the improvement achieved within approximately $20$ iterations. Although FPFA converges slightly more slowly than D-FPFA due to more connected antennas, both schemes have essentially converged by the $50^{\rm th}$ iteration, as further increasing the number of iterations to 200 improves the sum rate by less than 1\%. In our work, to balance performance and computational complexity, the maximum number of PSO iterations is set to $50$ in the subsequent simulations.

Next, we evaluate the sensitivity of the PSO-based MA optimization to the acceleration coefficients $c_1$ and $c_2$ in eq.~\eqref{eq:velocity_update} and the swarm size $N_{\rm par}$. As shown in Fig.~\ref{fig:c1c2}, when the value of $c_1$ and $c_2$ varies from $1.2$ to $2.2$, the sum rate changes by less than $1\%$. The selected setting $c_1=c_2=1.7$ achieves a sum rate close to that obtained with $c_1=c_2=2.2$. Similarly, Fig.~\ref{fig:swarmsize} shows that increasing the swarm size from $20$ to $60$ results in an overall rate variation of only about $1\%$. In particular, increasing the swarm size from the selected value of $40$ to $60$ improves the sum rate by less than $0.5\%$, while increasing the number of particles and the associated computational cost by $50\%$. Therefore, we set $c_1=c_2=1.7$ and swarm size as $40$ in the simulations. These results demonstrate that the adopted PSO algorithm is relatively insensitive to these key parameters within the considered ranges.

In Fig.~\ref{fig:SE_vs_gamma}, we evaluate the effect of the antenna-selection threshold $\gamma_{\mathrm{th}}$ on the sum rate. As $\gamma_{\mathrm{th}}$ increases from $0.6$ to $0.9$, more antenna connections are activated to retain a larger fraction of the coherent channel amplitude, resulting in a higher effective beamforming gain. However, when $\gamma_{\mathrm{th}}$ further increases to $1$, all candidate antenna connections are activated. Under the fixed total transmit-power constraint, the additional antennas with relatively weak channel gains provide limited useful contribution and may reduce the effective gain of the hybrid precoder after power normalization. Consequently, the sum rate first increases and then decreases with $\gamma_{\mathrm{th}}$. In the considered setting, the highest sum rate is obtained at $\gamma_{\mathrm{th}}=0.9$. However, the threshold that maximizes the sum rate may change with the user distribution and the resulting channel conditions. Therefore, $\gamma_{\mathrm{th}}=0.9$ should not be regarded as a universally optimal setting. Moreover, increasing $\gamma_{\mathrm{th}}$ activates more antenna connections and, correspondingly, more switches and phase shifters, leading to higher power consumption. The exact power consumption is further evaluated in Sec.~\ref{sec_SE_EE}. In comparison, $\gamma_{\mathrm{th}}=0.8$ achieves a sum rate within $0.5\%$ of the maximum while requiring fewer active hardware components. Therefore, to balance the sum rate and hardware power consumption, we set $\gamma_{\mathrm{th}}=0.8$ in the subsequent simulations.

Figs.~\ref{fig:SE_vs_movableregion} and \ref{fig:SE_vs_numMA} evaluate the effects of the movable-region size and the number of movable transmit subarrays, respectively. In Fig.~\ref{fig:SE_vs_movableregion}, we assume the BS contains $16$ movable subarrays and the side length of their square movable regions is varied from \(10\lambda_{\max}\) to \(50\lambda_{\max}\). The reference positions of adjacent subarrays are separated by \(50.5\lambda_{\max}\). Therefore, we limit the movable-region side length to \(50\lambda_{\max}\) to prevent the assigned regions from overlapping. In Fig.~\ref{fig:SE_vs_numMA}, when the number of movable subarrays is increased from $0$ to $16$, the remaining subarrays stay at their initial positions and the movable-region side length is fixed at $p = 10\lambda_{\max}$. The receiver positions are kept fixed in both evaluations to isolate the effect of transmit-subarray movement. The results show that the sum rate generally increases with both the number of movable subarrays and the movable-region size. This is because the additional spatial degrees of freedom allow the transmit-array geometry to better adapt to the channel conditions and mitigate inter-user interference. In practice, the number of subarrays and their movable regions are jointly constrained by the array aperture and movable-antenna hardware. In this work, the transmit array is divided into $16$ subarrays, and the movable-region side length is set to $10\lambda_{\max}$ to provide spatial position diversity while maintaining a moderate PSO searching space. Therefore, these parameters are adopted as an illustrative setting that balances the achievable rate improvement with the hardware and computational complexity.

\subsection{Sum rate and energy efficiency evaluation}
\label{sec_SE_EE}
\begin{table}
\centering
\caption{Configurations of the compared schemes.}
\label{Table_benchmarks}
\begin{tabular}{|c|c|c|c|c|}
\hline
\textbf{Scheme}& \textbf{\makecell{Tunable\\LO}} & \textbf{Movable} & \textbf{\makecell{Antenna \\ selection}} & \textbf{\makecell{Beam split\\multiplexing}}\\
\hline
D-FPFA & Yes & Yes & Yes  & Yes\\
D-FPFA + TTD & Yes & Yes & Yes & No\\
FPFA & Yes & Yes & No & Yes\\
D-FFA & Yes & No & Yes & Yes\\
FFA & Yes& No & No & Yes\\
Compact UPA + PS & No & No & No & No\\
Sparse UPA + PS & No & No & No & No\\
Sparse UPA + TTD & No & No & No & No\\
AoSA + PS & No & No & No & No\\
AoSA + TTD & No &No & No& No\\
\hline
\end{tabular}
\end{table}

\begin{table*}
\centering
\caption{Power consumption of the hardware components for the considered architectures.} \label{tab: power}
\begin{tabular}{|c|c|c|c|c|c|c|c|c|}
\hline
\multirow{2}*{Devices}& \multirow{2}*{\makecell[c]{Power \\ (mW)}}& \multicolumn{7}{c|}{Number of devices}\\
\cline{3-9}
 & & D-FPFA & D-FFA & FPFA& FFA & D-FPFA + TTD & \makecell{PS-based \\ fixed array} & \makecell{TTD-based \\ fixed array}\\
\hline
Baseband & 200 \cite{dynamic} & K & K & K & K&K &1 &1\\
\hline
RF chain with conventional LO& 120 \cite{power_RFchain}& 0 & 0 & 0 & 0 & 0 & U & U\\
\hline
RF chain with tunable LO & 200 \cite{power_tunable}& G & G & G & G & G &0 & 0\\
\hline
Phase shifter & 42 \cite{dynamic} & $\sum_{g=1}^GN_t^g$ &$\sum_{g=1}^GN_t^g$ & $GN_t$&$GN_t$ & 0 & $UN_t$ & 0\\
\hline
Switch & 24 \cite{PS_power} & $\sum_{g=1}^GN_t^g$ &$\sum_{g=1}^GN_t^g$ &0 & 0&$\sum_{g=1}^GN_t^g$ & 0 & 0\\
\hline
TTD & 80 \cite{EE_FTTD} & 0 & 0 &0 & 0 &$\sum_{g=1}^GN_t^g$ &0 & $\sum_{g=1}^GN_t^g$\\
\hline
Movable antenna driver & $3000$ \cite{MA_power} &$N_{\rm sub}$ & 0 & $N_{\rm sub}$ &$0$ & $N_{\rm sub}$ & 0 & 0\\
\hline
\end{tabular}
\end{table*}

\begin{figure} 
\centering 
\includegraphics[width = 0.4\textwidth]{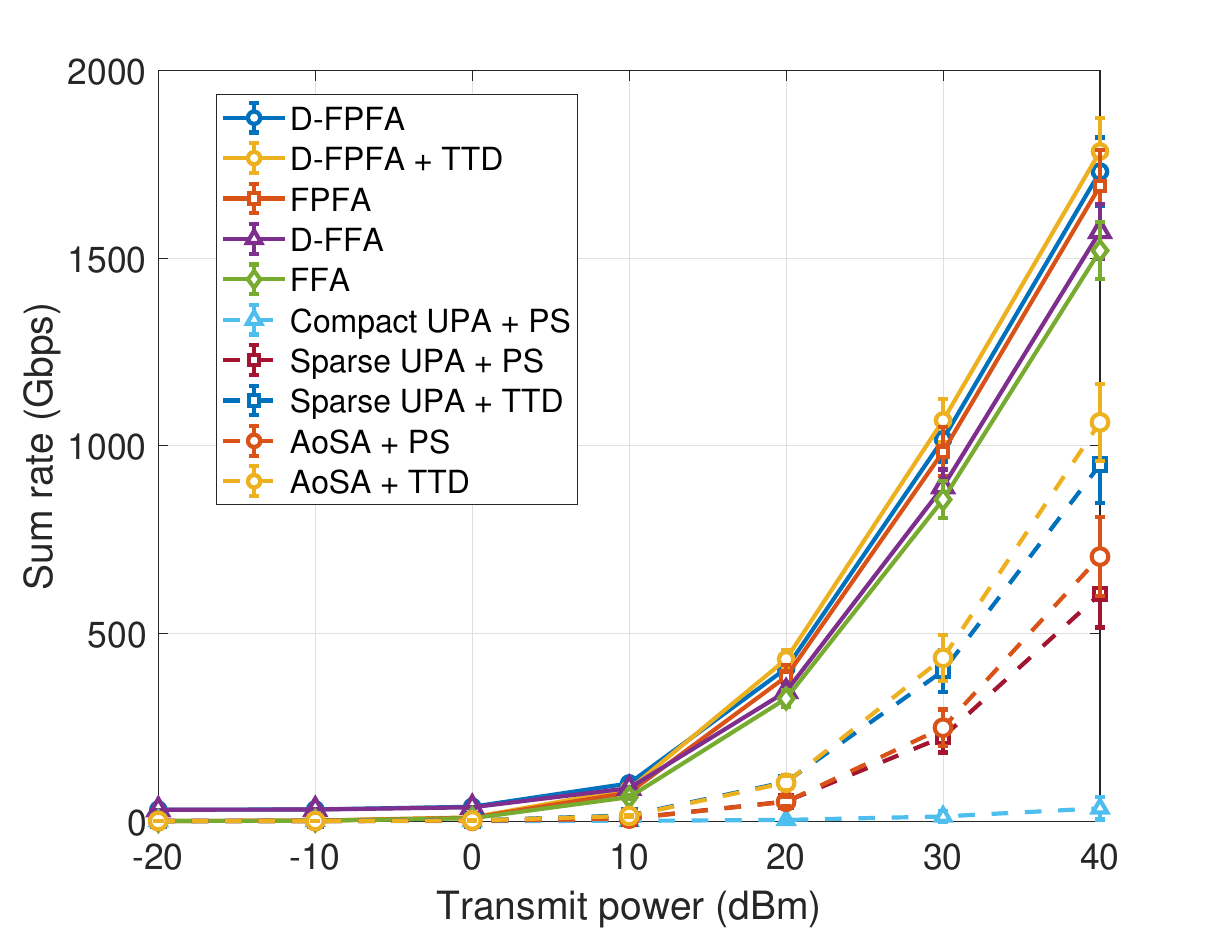} 
\caption{Sum rate versus transmit power.} 
\label{Fig_SE_vs_power} 
\end{figure}

\begin{figure} 
\centering 
\includegraphics[width = 0.4\textwidth]{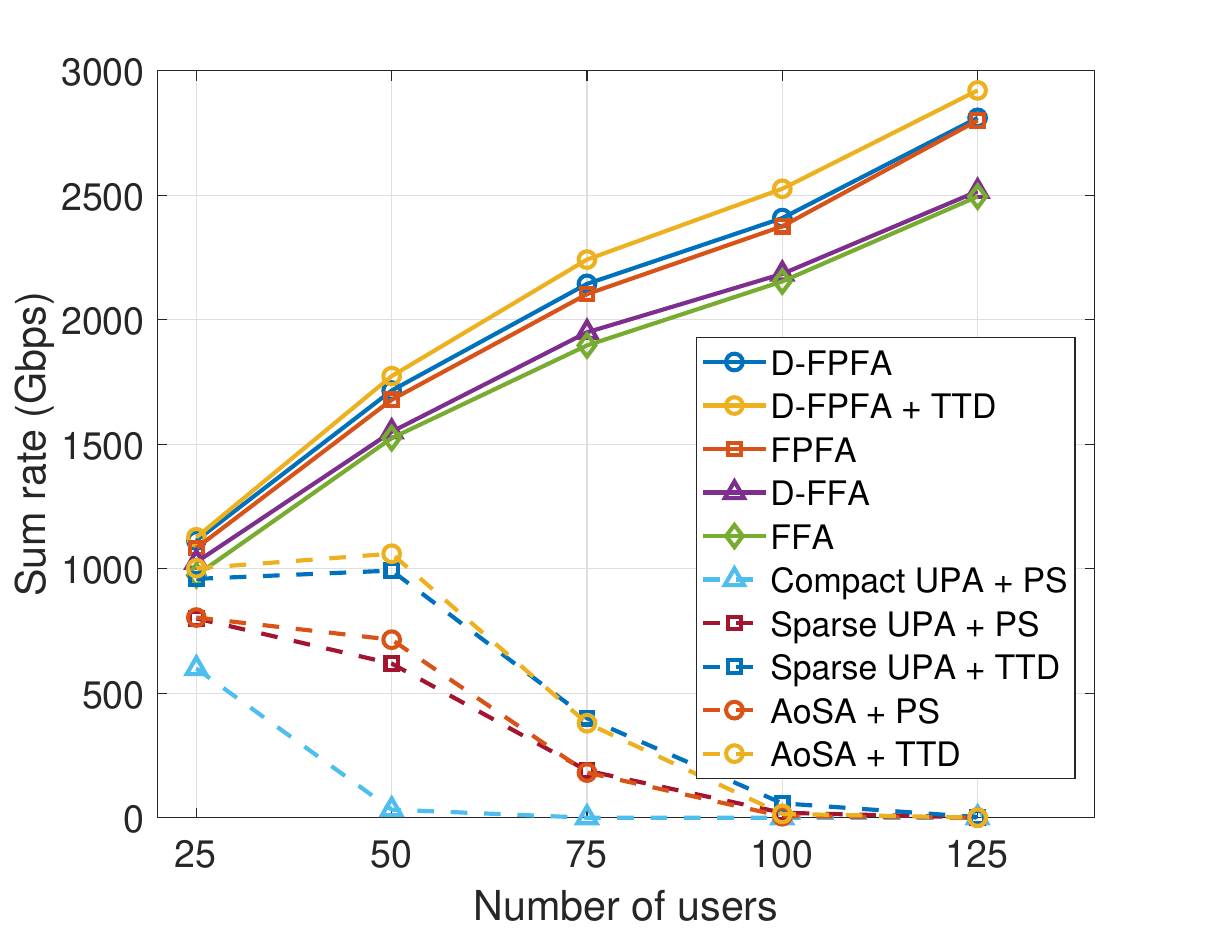} 
\caption{Sum rate versus number of users.} 
\label{Fig_SE_vs_UE} 
\end{figure}

In this section, we compare the proposed D-FPFA with two categories of benchmarks. The first category employs tunable LOs to access $K$ sub-bands. It includes four PS-based frequency-fluid architectures, namely D-FPFA, FPFA, D-FFA, and FFA. The four PS-based architectures differ in whether joint transceiver-position optimization and dynamic antenna selection are enabled. Moreover, a TTD-based D-FPFA benchmark is added. For the PS-based schemes, the analog focusing vectors are designed at the corresponding central frequency and remain fixed over the subcarriers so that the frequency-dependent beam-split effect is retained. In TTD-based schemes, we assume ideal TTD so that the frequency-dependent phase variation can be perfectly compensated, and the beam split effect is therefore suppressed. The second category consists of five frequency-fixed benchmarks without tunable LOs. These benchmarks operate at one central frequency over a 5-GHz band and employ different fixed array geometries with PS or ideal-TTDs. In this work, we control the array aperture of all the schemes to be approximately the same. The configurations of all compared schemes are summarized in Table~\ref{Table_benchmarks}.

In Fig.~\ref{Fig_SE_vs_power}, we evaluate the sum rate versus the transmit power. The markers report the average results over $1000$ independent Monte Carlo realizations with randomly generated user locations, while the error bars represent the standard deviation. As expected, the sum rate of all the considered architectures increases with the transmit power. The four PS-based tunable-LO-based architectures, i.e., D-FPFA, FPFA, D-FFA and FFA, consistently outperform the fixed-frequency benchmarks. This is because the proposed two-phase frequency allocation strategy separates users with strong potential interference across different sub-bands and further exploits the beam-split effect to serve multiple users through different subcarriers of one wideband beam. Specifically, at $40$~dBm, D-FPFA achieves approximately $2.3$ times the sum rate of the conventional fixed-position AoSA with PSs. Then, by comparing D-FPFA and FPFA with their fixed-position counterparts, D-FFA and FFA, respectively, we can observe the contribution of the movable-subarray positioning. In particular, D-FPFA provides an approximately 10\% sum rate improvement over D-FFA at $40$~dBm.
We next compare the PS-based architectures with their TTD-based counterparts, and observe that all TTD-based implementations achieve higher sum rate. This is because the PS-based analog beamforming vectors are designed at the central frequency and remain unchanged across the subcarriers, whereas ideal TTDs provide frequency-dependent phase compensation and thereby eliminate the beam-split effect. Specifically, at $40$~dBm, the AoSA with PSs attains only around 70\% of the sum rate achieved by its TTD counterpart, while D-FPFA reaches approximately 95\% of its corresponding TTD result. This smaller gap is mainly attributed to beam-split multiplexing. By assigning different subcarriers of one wideband beam to users aligned with the corresponding split beams, D-FPFA reduces intra-group interference, so that the performance loss caused by the wideband beam split effect can be compensated.
Finally, we analyze the dynamic antenna-selection mechanism by comparing D-FPFA and D-FFA with FPFA and FFA, respectively. We observe that the dynamic architectures achieve a slightly higher sum rate. This observation is consistent with the previous discussion in Sec.~\ref{sec_antenna_selection} that the dynamic antenna-selection mechanism can increase sum rate by deactivating weak antenna connections.

\begin{figure}
    \centering
    \includegraphics[width=0.4\textwidth]{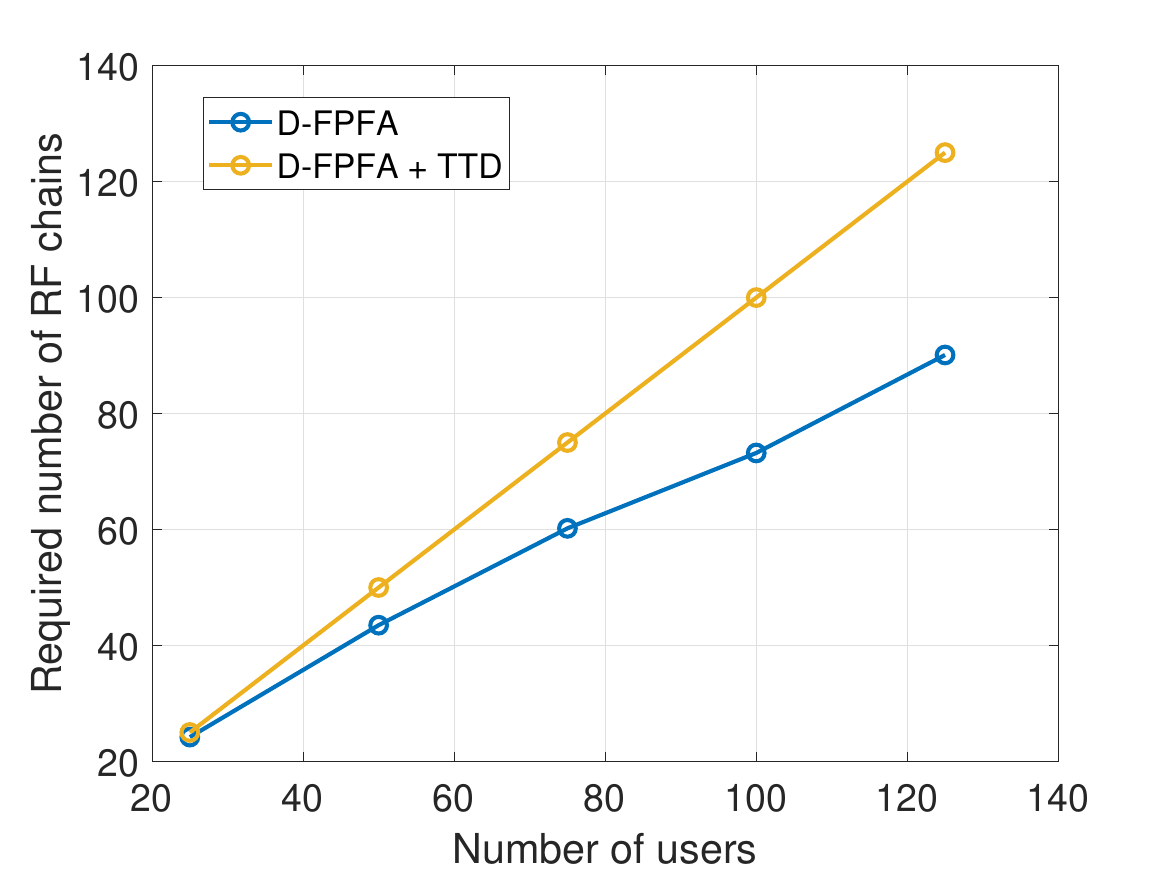}
    \caption{Required RF chain count versus the number of users.}
    \label{fig:RF_chain_count}
\end{figure}

\begin{figure} 
\centering 
\includegraphics[width = 0.45\textwidth]{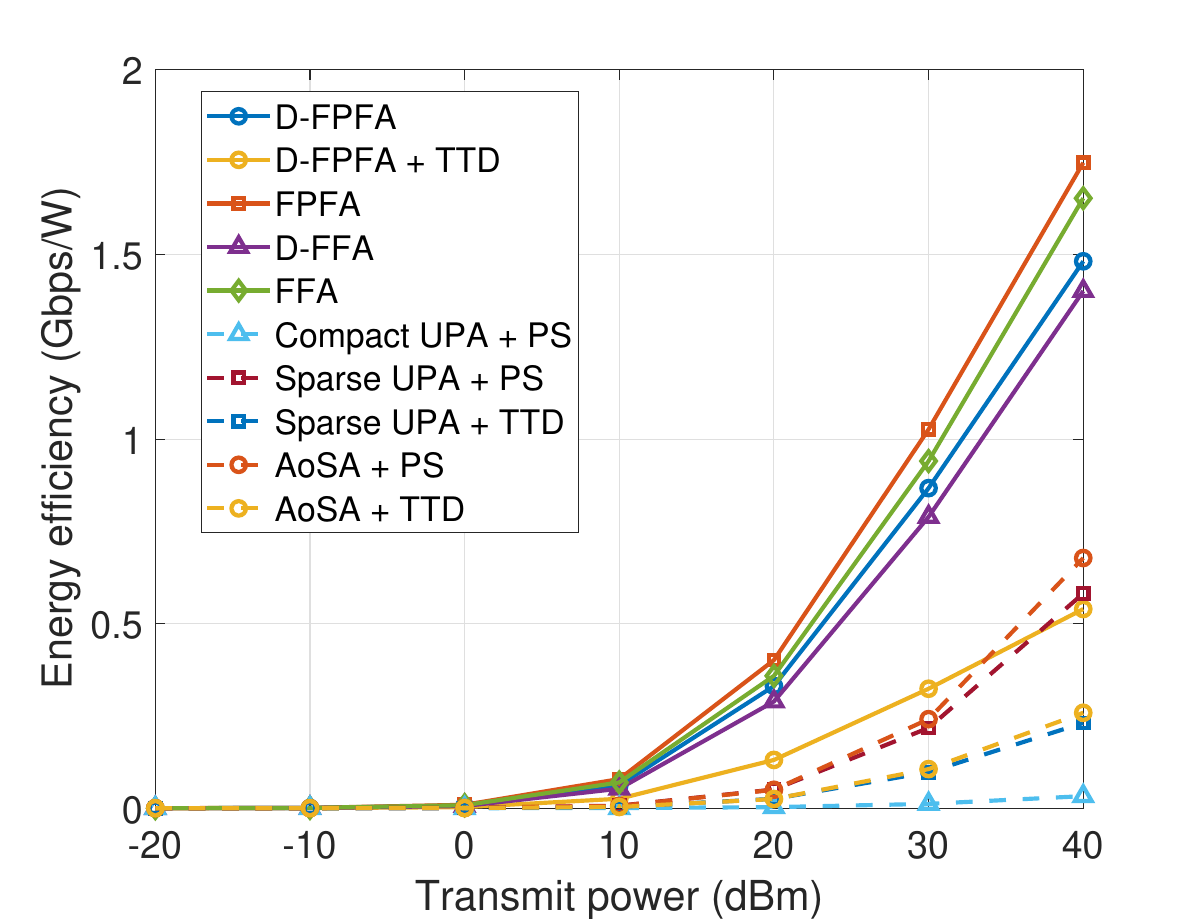} 
\caption{Energy efficiency versus transmit power.} 
\label{Fig_EE_vs_power} 
\end{figure}

Fig.~\ref{Fig_SE_vs_UE} presents the sum rate versus the number of users. We observe that the proposed tunable LO-based architectures maintain a steady increase in sum rate as the user population grows since they utilize the frequency domain multiplexing. In contrast, the conventional fixed-frequency benchmarks suffer substantial performance degradation in the high-load regime as the spatial multiplexing alone fails to distinguish these users simultaneously.
Among all the considered architectures, D-FPFA with ideal TTDs consistently achieves the highest sum rate because it eliminates the beam split effect. However, this performance gain is achieved at the cost of increased hardware cost. On one hand, the TTDs consume more power than conventional PSs. On the other hand, as the TTD-based benchmark suppresses rather than exploits beam split, it requires at least one RF chain per user. In contrast, D-FPFA can serve multiple grouped users through different subcarriers of one wideband beam, thus reducing the required number of RF chains to the number of user groups $G$.
To quantify this additional hardware requirement, Fig.~\ref{fig:RF_chain_count} compares the required number of RF chains in D-FPFA and D-FPFA with TTD. It can be observed that the RF-chain count of the TTD-based architecture increases directly with the number of users, while D-FPFA requires much fewer RF chains. These results demonstrate the trade-off between higher sum rate provided by TTDs and lower hardware complexity and power consumption enabled by the proposed beam-split multiplexing.

In Fig.~\ref{Fig_EE_vs_power}, we compare the energy efficiency (EE) for different architectures, which is defined as the ratio of the sum rate to the total power consumption. The power consumption of each hardware component and the device counts are summarized in Table~\ref{tab: power}, where $N_t^g$ denotes the number of antennas connected to the $g_{\rm th}$ RF chain. For a fixed-frequency RF chain, the total power consumption is $120$ mW, including $5$~mW for the conventional LO~\cite{power_RFchain}. The tunable LO considered in~\cite{power_tunable} consumes $51.7$ to $84.1$~mW. Using the conservative upper-bound value, the power consumption of a tunable-LO-based RF chain is therefore set to $P_{\mathrm{RF,tunable}}=120-5+84.1\approx 200~\mathrm{mW}$.
From the figure, we observe that FPFA achieves the highest EE over the evaluated transmit-power range among all the considered schemes. Specifically, at $40$~dBm, FPFA reaches approximately $1.75$~Gbps/W, which is about $2.5$ times that of the AoSA with PSs. At the same transmit power, D-FPFA achieves an EE of approximately $1.48$~Gbps/W, corresponding to about $2$ times that of the same conventional benchmark. These results show that both proposed architectures provide substantial EE improvements, with FPFA achieving the highest EE and D-FPFA attaining higher sum rate while maintaining competitive EE. We next examine the effects of TTDs and dynamic antenna selection. Although as illustrated in Fig.~\ref{Fig_SE_vs_power}, ideal TTDs provide higher sum rate by eliminating the beam-split effect, all the TTD-based architectures exhibit lower EE than their PS-based counterparts. This is because TTDs consume considerably more circuit power than PSs and, without beam-split-based user grouping, require more RF chains to serve the same number of users. In particular, the PS-based D-FPFA achieves approximately $2.8$ times the EE of its TTD counterpart at $40$~dBm, even though the latter attains the highest sum rate. The TTD counterpart also exhibits lower EE than the PS-based sparse UPA and AoSA benchmarks. These results indicate that, the sum rate improvement provided by the ideal TTDs cannot compensate for their additional hardware power, making them less suitable for energy-efficient implementations.
Moreover, we observe that D-FPFA and D-FFA achieve lower EE than FPFA and FFA, respectively. Although dynamic antenna selection provides a moderate sum rate improvement as discussed previously, this gain is insufficient to compensate for the power consumed by the switching network. In contrast, FPFA achieves slightly higher EE than FFA, suggesting that, for the considered number of movable subarrays and driver-power setting, the sum rate improvement obtained from movable-position optimization can compensate for the associated movement power. Overall, D-FPFA achieves the highest sum rate among the practical PS-based architectures while maintaining competitive EE, whereas FPFA achieves the highest EE and provides the best balance between system performance and hardware power consumption.

\subsection{Discussion and Limitations}
\label{sec_discussion}
Despite the advantages of the proposed architecture, several practical limitations should be noted. First, the proposed D-FPFA framework is developed under the assumption of perfect CSI. Therefore, the reported results should be interpreted as a performance upper bound under ideal CSI rather than as a robustness assessment under imperfect CSI. In practice, CSI errors may affect the frequency allocation, antenna-position optimization, and hybrid precoding procedures, thereby reducing the resulting sum rate. Moreover, acquiring the channel responses over multiple sub-bands, subcarriers, and candidate antenna positions may introduce considerable training and feedback overhead. 
Robust optimization and low-overhead channel estimation methods for movable-antenna systems have been investigated in~\cite{MA_Globally_optimal} and~\cite{MA-estimation}, respectively. These studies provide useful starting points for addressing the above limitations. However, they are based on the narrowband channel models, and their extension to the considered wideband D-FPFA system requires further investigation.
Second, the channel model considered in this work focuses on the LoS scenario, which provides a baseline for evaluating the proposed architecture in THz near-field communications. However, the present analysis does not quantify the effects of multipath propagation, blockage, or user mobility. Multipath components may affect the frequency-dependent beamforming gains and distort the beam-split coverage, thereby changing the user-grouping graph. Blockage may remove the dominant LoS component, while mobility introduces time-varying propagation distances and Doppler effects. These factors may affect both the channel characteristics and the resulting system design. Therefore, extending the proposed framework to more general propagation conditions and dynamic environments will be investigated in future work.

\section{Conclusion}
\label{section_conclusion}
In this paper, we proposed the D-FPFA architecture to support ultra-dense connectivity in THz massive MIMO systems. Based on this architecture, we first establish the near-field channel model and the system model. Then, we propose the two-phase frequency allocation algorithm to support massive connections. In the first phase, we allocate the sub-bands at the aim of minimizing the inter-user interference where we apply the channel correlation coefficients as the measure. In the second phase, we investigated the wideband near-field beam-split effect for planar arrays and showed that the beam at a non-central subcarrier generally cannot be perfectly refocused at a single spatial point. 
Accordingly, we characterized the beam-split coverage through the beamforming gain and formulated user grouping as a directed MDS problem. This formulation enables users within each group to be served by different subcarriers of a single wideband beam. To maximize the system sum rate from the spatial domain design, we developed a joint optimization framework. Leveraging the switching network, we first implemented a distance-based antenna selection to mitigate the near-field gain variations. Subsequently, a PSO-based algorithm was proposed to jointly optimize the positions of the movable arrays and the precoders. We evaluated the performance of the proposed architecture over different configurations. The numerical results demonstrated that the proposed tunable-LO-based architectures maintain increasing sum rate as the number of users grows, whereas the fixed-frequency benchmarks experience substantial degradation in the high-load regime. At $40$ dBm, the proposed D-FPFA achieved approximately $2.3$ times the sum rate of the conventional PS-based AoSA architecture. It also attained 95\% of the sum rate of its TTD counterpart while providing approximately $2.8$ times the EE of that counterpart. Moreover, FPFA, the fully connected variant of D-FPFA, achieved the highest EE among all considered architectures, reaching about $2.5$ times the EE of the conventional PS-based AoSA. Overall, D-FPFA provided the highest sum rate among practical PS-based designs and maintained competitive EE, whereas FPFA offered the best rate-power balance. Extending the proposed framework to imperfect CSI, more general propagation conditions, and user mobility remains an important directions for future work.

\section*{Appendix}
Consider a two-dimensional planar array located in the x-z plane, with $N = N_x \times N_z$ antennas. The coordinate of the $n$-th antenna element is denoted as $\mathbf{p}_t^n = [x_n, 0, z_n]^T$. We consider a target user with spherical coordinates $(r,\theta,\phi)$ and Cartesian position $\mathbf p_u=[u_x,u_y,u_z]$. Applying the analog beam focusing vector \eqref{eq_beam_focusing}, the ideal phase shift for the $n_{\rm th}$ antenna element to focus on this user at the central frequency $f_c$ is determined by the transmission distance $d_n$, which can be written as
\begin{equation}
    d_n = \sqrt{(u_x - x_n)^2 + u_y^2 + (u_z - z_n)^2},
\end{equation}
where
\begin{equation}
        u_x = r \sin\phi \cos\theta, u_y = r \sin\phi \sin\theta, u_z = r \cos\phi.
\end{equation}
For the NF SWM model, we apply the second-order Taylor expansion to $d_n$. By factoring out $r$ and retaining terms up to the quadratic order of the antenna coordinates $(x_n, z_n)$, the distance is approximated as
\begin{equation}
    d_n \approx r - \Phi_{\text{linear}}(x_n, z_n) + \Phi_{\text{quad}}(x_n, z_n),
\end{equation}
where
\begin{subequations}
    \begin{align}
        & \Phi_{\text{linear}} = x_n \sin\phi \cos\theta + z_n \cos\phi, \\
        & \Phi_{\text{quad}} = \frac{1}{2r} \left[ x_n^2 (1 - \sin^2\phi \cos^2\theta) + z_n^2 (1 - \cos^2\phi) \right.\\
        & \quad \quad \quad \quad \quad \quad \quad \left.- 2x_n z_n (\sin\phi \cos\theta \cos\phi) \right]\notag.
    \end{align}
\end{subequations}
Perfect phase matching at $f_m$ requires the phase residual to be independent of the antenna index, i.e.,
\begin{equation}
\label{eq_residual}
  f_m d_n(r_m,\theta_m,\phi_m)-f_c d_n(r,\theta,\phi)=C,\qquad \forall n,
\end{equation}
where $C$ is an antenna-independent constant and therefore does not affect coherent focusing. Consequently, the coefficients of $x_n$, $z_n$, $x_n^2$, $z_n^2$, and $x_nz_n$ must be matched separately.

First, matching the linear coefficients of \(x_n\) and \(z_n\) gives
\begin{subequations}
\begin{align}
  f_m a_m&=f_c a,\\
  f_m b_m&=f_c b,
\end{align}
where $a=\sin\phi\cos\theta,\qquad b=\cos\phi$, and $a_m=\sin\phi_m\cos\theta_m,\qquad b_m=\cos\phi_m$.
\end{subequations}
Let \(\eta=f_c/f_m\). The candidate drifted direction therefore satisfies \(a_m=\eta a\) and \(b_m=\eta b\), or equivalently,
\begin{subequations}
\begin{align}
  \phi_m&=\arccos(\eta\cos\phi),\\
  \theta_m&=\arccos\!\left(\frac{\eta\sin\phi\cos\theta}{\sqrt{1-\eta^2\cos^2\phi}}\right).
\end{align}
\end{subequations}

Second, matching the coefficients of \(x_n^2\) and \(z_n^2\) independently requires the candidate focal range \(r_m\) to satisfy
\begin{subequations}
\begin{align}
  r_{m,x}&=\frac{r}{\eta}\frac{1-\eta^2a^2}{1-a^2},\\
  r_{m,z}&=\frac{r}{\eta}\frac{1-\eta^2b^2}{1-b^2}.
\end{align}
\end{subequations}
Their difference is
\begin{equation}
  r_{m,x}-r_{m,z}
  =\frac{r}{\eta}\frac{(1-\eta^2)(a^2-b^2)}{(1-a^2)(1-b^2)}.
\end{equation}%
Therefore, for a nonzero frequency offset $\eta\neq1$ and a general direction satisfying $a^2\neq b^2$, the two aperture dimensions require different candidate focal ranges, i.e., \(r_{m,x}\neq r_{m,z}\). Hence, no single \(r_m\) can simultaneously match the \(x_n^2\)- and \(z_n^2\)-dependent phase-curvature terms. Moreover, when \(ab\neq0\), matching the mixed \(x_nz_n\) coefficient imposes the additional requirement as
\begin{equation}
  r_m=r_{m,xz}=\eta r,
\end{equation}
which further restricts the existence of a common drifted focal point.
Consequently, at \(f_m\neq f_c\), the residual phase in (\ref{eq_residual}) cannot generally be made independent of the antenna index over both aperture dimensions. Thus, the phases of all antenna contributions cannot be perfectly matched at one common spatial point, and the beam cannot be perfectly refocused there. This proves the dimension-dependent focal mismatch, i.e., the astigmatism phenomenon stated in Theorem 1. $\hfill \blacksquare$
\bibliographystyle{IEEEtran}
\bibliography{arXiv}
\end{document}